\documentstyle[dina4,12pt]{article}

\begin{document}

\newcommand{\be}{\begin{equation}}
\newcommand{\ee}{\end{equation}}
\newcommand{\bea}{\begin{eqnarray}}
\newcommand{\eea}{\end{eqnarray}}
\newcommand{\no}{\nonumber \\}
\newcommand{\fs}{\; \; .}
\newcommand{\co}{\; \; ,}
\newcommand{\eff}{e\hspace{-0.1em}f\hspace{-0.18em}f}

\begin{titlepage}
\begin{flushright}BUTP-93/24\end{flushright}
\rule{0em}{2em}\vspace{4em}
\begin{center}
{\LARGE {\bf On the foundations of chiral perturbation theory}}\\ \vspace{2em}
H.Leutwyler\\Institut f\"{u}r theoretische Physik der Universit\"{a}t Bern\\
Sidlerstr. 5, CH-3012 Bern, Switzerland\\
\vspace{2em}
August 1993\\
\vspace{5em}
{\bf Abstract} \\
\vspace{2em}
\parbox{30em}{
The properties of the effective field theory relevant for the low energy
structure generated by the Goldstone bosons of a spontaneously broken
symmetry are reexamined.
It is shown that anomaly free, Lorentz
invariant theories are characterized by a gauge invariant effective Lagrangian,
to all orders of the low energy expansion. The paper includes
a discussion of anomalies and approximate
symmetries, but
does not cover nonrelativistic effective theories.} \\
\vspace{4em}
\rule{30em}{.02em}\\
{\footnotesize Work
supported in part by Schweizerischer Nationalfonds}
\end{center}
\end{titlepage}

\section{Introduction}
\label{intro}
The pioneering work on chiral perturbation theory was based on global symmetry
considerations \cite{pioneer,Callan,Dashen,Pagels}. The key
observation,
which gave birth to this development, is that a suitable effective field
theory involving Goldstone fields automatically generates transition
amplitudes which obey the low energy theorems of current
algebra and PCAC. The interaction among the Goldstone bosons is described by
an effective Lagrangian, which is invariant under global chiral
transformations. The insight
gained thereby not only led to a considerable simplification of current algebra
calculations, but also paved the way to a systematic investigation of the low
energy structure \cite{Weinberg1979,GLAnnals,Ungarn}.

The line of reasoning used to determine the form of the
effective theory, however, is of heuristic nature --- a compelling
analysis, which derives the properties of the effective Lagrangian from those
of the underlying theory,
is still lacking.
The problem with the standard "derivation" is that it
is based on {\it global}$\,$ symmetry considerations. Global symmetry
provides important
constraints, but does not suffice to determine the low energy structure.
A conclusive framework only results if
the properties of the theory are analyzed off the mass shell: one needs to
consider Green functions and study the
Ward identities which express the symmetries of the underlying theory at
the {\it local} level.

The
occurrence of anomalies illustrates the problem: massless QCD is
invariant under global $\mbox{SU}(\mbox{N}_f)_{\mbox{\scriptsize R}}
\times \mbox{SU}(\mbox{N}_f)_{\mbox{\scriptsize L}} $, but, unless $\mbox{N}_f
\leq 2$, the corresponding effective Lagrangian is not.
Indeed, it is well-known from Noether's theorem that a noninvariant
Lagrangian may describe a symmetric theory: under the action of the
symmetry group, the Lagrangian may only pick
up a total derivative, such that the action is not affected.
This is precisely what happens in the presence of
anomalies. Moreover, a similar phenomenon also occurs in nonrelativistic
effective theories. The effective Lagrangian of a ferromagnet, e.g.,
is invariant under rotations of the spin directions
only up to a total derivative \cite{Ferromagnet}.

These examples indicate that there is a loophole in the heuristic argument:
it is not legitimate to postulate that the effective theory is
characterized by a symmetric Lagrangian.
Instead, the low energy analysis should exclusively rely on the Ward identities
of
the underlying theory and the properties of the effective Lagrangian
should be derived from there. The purpose of the present paper is to
show that this can indeed be done. The result demonstrates
that the "Current algebra plus PCAC" technique is strictly
equivalent to the effective Lagrangian method.
More specifically, it will be shown that, if
the underlying theory is Lorentz invariant and does not contain anomalies,
then the Ward identities insure that the low energy structure of the Green
functions may be described in terms of an effective field theory
with a symmetric effective Lagrangian.

\section{Generating functional, Ward identities}
\label{gen} \setcounter{equation}{0}
Consider a spontaneously broken exact symmetry\footnote{The
analysis is extended to approximate
symmetries in section \ref{appr}.}: the Hamiltonian of the theory is invariant
under a Lie group G, but the ground state is
invariant only under the subgroup $\mbox{H}\subset\mbox{G}$. For each one of
the generators of G, there is a conserved current $J^\mu_ i(x),\;i=1,\ldots,
d_{\mbox{\scriptsize G}}$.
Assuming that
the spontaneous symmetry breakdown gives rise to order parameters ---
vacuum expectation values of local operators with nontrivial
transformation properties under G --- the Goldstone theorem
\cite{Goldstone} then asserts
that (i) the spectrum of the theory contains $\mbox{N}_{\mbox{\scriptsize GB}}=
d_{\mbox{\scriptsize G}}-d_{\mbox{\scriptsize H}}$ massless particles
($d_{\mbox{\scriptsize G}}$ and $d_{\mbox{\scriptsize H}}$ count
the generators of G and H, respectively) and (ii) the transition
matrix elements of the currents between the vacuum and the Goldstone bosons
are different from zero. Using QCD-terminology, I refer to
these particles as "pions", denoting the corresponding one-particle states by
$\mid\!\pi^a(p)\!>$. The index $a=1,\ldots, \mbox{N}_{\mbox{\scriptsize
GB}}\,$ labels the different Goldstone flavours and $p$ is the
four-momentum. Lorentz invariance implies that the transition matrix elements
are of the form\be\label{found1}
<\!0\!\mid J^\mu_i\mid\!\pi^a(p)\!> =i F^a_i p^\mu\fs\ee
According to the Goldstone theorem, the $d_{\mbox{\scriptsize
G}}\times\mbox{N}_{\mbox{\scriptsize GB}}$ matrix
$F^a_i$ is of rank $\mbox{N}_{\mbox{\scriptsize GB}}$.
In a suitable
basis, the states $\mid\!\pi^a(p)\!>$ are orthogonal and the "decay constants"
$F^a_i$ are real and diagonal, $F^a_i=\delta^a_i F_{(i)}$; in particular,
these constants vanish if the index $i$ labels one of the currents of
the subgroup H. The matrix $F^a_i$ may contain several independent
eigenvalues \cite{Boulware}. The number of independent eigenvalues depends on
the structure of the Lie algebra {\bf G} of the group.\footnote{Discrete
symmetries may yield additional constraints.}
Denote the subalgebra
spanned by the generators of H by {\bf H} and set {\bf G}={\bf H}+{\bf K}.
The subspace {\bf K} carries a representation $D_{\mbox{\scriptsize {\bf
K}}}(h)$ of the subgroup H. If this representation
is irreducible, then there is a single decay
constant. Otherwise, the number of independent eigenvalues of the
matrix $F^a_i$ is given by the number of irreducible components of
the representation $D_{\mbox{\scriptsize {\bf K}}}(h)$. Since the vectors of
the subspace {\bf K} are in one-to-one correspondence with the Goldstone
bosons, the various components represent pion multiplets,
transforming irreducibly under H.

The following analysis deals with the Green functions
formed with the currents.
It is convenient to collect these in the generating functional $\Gamma\{f\}$,
defined by\bea\label{found2}
e^{i\,\Gamma\{f\}}=\sum_{n=0}^\infty\frac{i^n}{n!}\int\!d^d\!x_1\ldots
d^d\!x_n \;f_{\mu_1}^{i_1}(x_1)\ldots f_{\mu_n}^{i_n}(x_n)\times\no
\times <\!0\!\mid T\{J^{\mu_1}_{i_1}(x_1)\ldots
J^{\mu_n}_{i_n}(x_n)\}\mid\!0\!>\co\eea
where $f_\mu^i(x)$ is a set of external fields, which play the role of
auxiliary variables.
The generating functional admits a simple intuitive interpretation.
The external field may be viewed as a modification of the Lagrangian:
${\cal L}\rightarrow{\cal L}+f_\mu^iJ^\mu_i$. Suppose that, in
the remote past, the system was in the ground state and consider the evolution
in the presence of the external field. The quantity
$e^{i\,\Gamma\{f\}}$ is the vacuum-to-vacuum transition amplitude, i.e.,
represents the probability amplitude
for the system to wind up in the ground state when $x^0\rightarrow +\infty$.

In the language of the generating functional, the Ward identities obeyed by the
Green
functions of the currents take a remarkably simple form: in the absence of
anomalies, the Ward identities are equivalent to the statement that the
generating functional is invariant under gauge transformations of the
external
fields,
\be\label{found3}\Gamma\{T(g)f\}=\Gamma\{f\}\fs\ee
Gauge transformations correspond to space-time-dependent group
elements, i.e., to a map from Minkowski space into the group,
$x\rightarrow g(x)\in
\mbox{G}$. To specify the action of the group on the external fields, it
is convenient to use a matrix representation for these fields. Consider a
representation $D(g)$ of the group. The generators
$t_i$ of this representation obey the commutation
relation\be\label{found4} [t_i,t_j]=if^k_{\;ij}t_k\co\ee
where the $f^k_{\;ij}$ are the structure constants of the group.
The corresponding matrix representation of the
external fields is defined in terms of the generators as
$f_\mu(x)=\sum_it_if_\mu^i(x)$. In this notation,
the external fields transform according to\be\label{found5}
T(g)f_\mu(x)\equiv D(x)f_\mu(x)D^{-1}(x)
-i\partial_\mu D(x) D^{-1}(x)\ee
with $D(x)\equiv D\{g(x)\}$.
In particular, the change in the external fields generated by an
infinitesimal gauge transformation is given
by\be\label{found6}
\delta f_\mu^i(x)=\partial_\mu g^i(x) +f^i_{\;jk}f_\mu^j(x)g^k(x)\co\ee
where $g^1(x), g^2(x),\ldots$ are the infinitesimal coordinates of the group
element.

The invariance of the generating functional under gauge transformations of the
external fields expresses
the symmetry properties of the theory on the level of the Green functions.
It represents the basic ingredient of the
following analysis, while the specific properties, which the theory may
otherwise have, do not play any role.

Note that the generating functional is gauge invariant only if the Ward
identities obeyed by the Green functions of the currents do not contain
anomalies. If anomalies do occur, the generating functional transforms in a
nontrivial manner under the group. Throughout the first part of
this paper, anomalies are disregarded, such that equation
(\ref{found3}) is valid as it stands. The modifications
required to account for anomalous Ward identities are discussed in section
\ref{anom}.

\section{Pion pole dominance}
\label{ppd}\setcounter{equation}{0}

If the spectrum of asymptotic states contains a mass gap, the low energy
structure is trivial: the Fourier transforms of the Green
functions admit a straightforward Taylor series expansion in powers of the
momenta.
The low energy structure of theories with a spontaneously broken symmetry is
nontrivial, because the spectrum contains massless particles --- the
singularities generated by the exchange of Goldstone bosons do not admit
a Taylor series expansion in powers of the momentum. The two-point-function of
the current, e.g., contains a pole term due to the exchange of a pion,
\be\label{ppd1}
\int \!d^d\!x e^{ipx}<\!0\!\mid T\{J^\mu_i(x) J^\nu_k(0)\}\mid\!0\!>=
i\frac{p^\mu p^\nu}{p^2 +i\epsilon}\sum_a F^a_iF^a_k +\ldots\ee
The current algebra analysis
of the low energy structure
is based on the assumption that
\begin{enumerate}
\item The Goldstone bosons generated by spontaneous
symmetry breakdown are the only massless particles contained in the
spectrum of asymptotic states.
\item At low energies, the Green functions are dominated by the poles due to
the exchange of these particles.
\end{enumerate}
The pole terms represent one-particle-reducible contributions, involving
the propagation of a pion between the various vertices. In the case of the
two-point-function, the pole term corresponds to a graph where the first
current emits a pion which propagates and gets absorbed by the second one.
The three-point-function may receive contributions associated with the exchange
of one, two or three pions: Graphs involving a single pole connect two
vertices, one of which involves the coupling of a pion line to one
current, while the other represents an interaction between a pion and two
currents;
three poles occur if each of the three currents emits a pion, propagating to a
vertex, where the three pions interact with one another, etc. The cluster
decomposition property insures that the vertices occurring in the various
graphs are independent of the particular Green function under
consideration --- they only depend on the momenta and the flavour quantum
numbers of the pions and currents which enter the vertex in question.

Clustering implies
that the simultaneous exchange of more than one pion between the
same two vertices necessarily also occurs. These
processes may be pictured as graphs containing loops. Again, the vertices
occurring therein are the same as for the one-particle-reducible
terms, i.e., for the tree graphs. The corresponding
contribution to the Green function contains a cut in the relevant momentum
transfer rather than a pole.

The third assumption entering the "current algebra plus PCAC" analysis is that
\begin{enumerate}\setcounter{enumi}{2}
\item The vertices admit a Taylor series expansion in powers of the
momenta.\end{enumerate}
The vertices determine the residues of the
poles due to one-pion-exchange --- the assumption amounts to the
hypothesis that these residues may be expanded in a Taylor series.
This represents a quantitative formulation of the first
assumption: the only singularities occurring at low energies are the poles
and cuts due to the exchange of Goldstone bosons. Once these singularities are
accounted for, the amplitudes do admit a Taylor series expansion.

An immediate consequence is that, at low
energies, the hidden symmetry prevents the Goldstone bosons from
interacting with one another. Since this property is essential for the
consistency of the low
energy analysis to be described in the remainder of this paper, I
briefly review the argument \cite{Weinberg1966}.
Consider the probability amplitude for the currents to create pions
out of the vacuum. Current conservation requires
\be
\label{eft30}
p_\mu<\!\pi^{a_1}(p_1)\pi^{a_2}(p_2)\ldots \;\mbox{out}\mid\!
J^\mu_i\!\mid\!0\!>=0\co \ee
where $p^\mu= p^\mu_1+p^\mu_2+\ldots $ is the four-momentum of the final state.
The amplitude for pair creation, e.g., may contain a pole term
proportional to the three-pion vertex $v_{a_1a_2a_3}(p_1,p_2,p_3)$,
\[<\!\pi^{a_1}(p_1)\pi^{a_2}(p_2) \;\mbox{out}\mid \!J^\mu_i\!\mid\!0\!>
=\,-i\,
\frac{p_3^\mu}{p_3^2+i\epsilon}\sum_{a_3}F^{a_3}_i\;v_{a_1a_2a_3}(p_1,p_2,p_3)+
\ldots\]
As this
represents the only one-particle-reducible contribution to the amplitude in
question, the remainder is free of poles. According to 2., the pole
term dominates the amplitude at low energies, while 3. implies that
the vertex admits a
Taylor series expansion, starting with a momentum independent term
$v_{a_1a_2a_3}(0,0,0)$. In the limit $p^\mu_1,p^\mu_2,p^\mu_3\rightarrow 0$,
current conservation thus yields
$\sum_{a_3}F^{a_3}_iv_{a_1a_2a_3}(0,0,0)=0$. In view of the rank of the
matrix
$F^a_i$, this
implies that the three-pion-vertex vanishes in the zero momentum limit,
$v_{a_1a_2a_3}(0,0,0)=0$.

The
one-particle-reducible pieces can unambigously be distinguished from the
remainder only through the singularities which they produce --- the residues of
the poles may be evaluated on the mass shell of the particles which meet at the
vertex in question.
For the three-pion-vertex, this means that only the
value at $p_1^2=p_2^2=p_3^2=0$ is of physical significance. On account of
Lorentz invariance and momentum conservation, the function
$v_{a_1a_2a_3}(p_1,p_2,p_3)$ only
depends on the three scalars $p_1^2,p_2^2$ and $p_3^2$. Since the vertex
vanishes for $p_1=p_2=p_3=0$, it vanishes everywhere on the
mass shell of the three particles, i.e.,
one-particle-reducible contributions
containing a triple pion vertex do not occur, in any amplitude. Note that
Lorentz invariance plays an essential role here --- in the
nonrelativistic regime, kinematics does not prevent
three Goldstone bosons from interacting with one another.

The production amplitude for three pions contains at most a
single pole from a tree graph, which involves a four-pion vertex joined to the
current by a single pion propagator,
\[<\!\pi^{a_1}(p_1)\pi^{a_2}(p_2)\pi^{a_3}(p_3) \;\mbox{out}\mid\!
J^\mu_i\!\mid\!0\!> =-i
\frac{p^\mu_4}{p^2_4+i\epsilon}\sum_{a_4}F^{a_4}_i\;v_{a_1a_2a_3a_4}
(p_1,p_2,p_3,p_4) + \ldots\]
At low momenta, this term again dominates over the remainder,
such that current conservation implies $v_{a_1a_2a_3a_4}(0,0,0,0)=0$:
the hidden symmetry prevents pions of zero momentum from
scattering elastically. In contrast to the preceding case,
the four scalars
$p_1^2,\ldots,p_4^2$ are, however, not the only Lorentz invariants which can be
formed with the momenta: the Mandelstam variables are
independent thereof. The residue of the
pole thus becomes a function of say, $s$ and $t$, and the low energy expansion
starts with
$v_{a_1\ldots a_4}(p_1,\ldots,p_4)=c^1_{a_1\ldots a_4}\,s+
c^2_{a_1\ldots a_4}\,t+O(p^4).$

Using induction, the argument readily extends to vertices with arbitrarily
many pion legs: if the interaction among up to $n$ pions is suppressed by two
powers of momentum, the relation (\ref{eft30}) implies that this is the
case also for $n\!+\!1$ pions: independently of the number of
pions participating in the interaction, the scattering amplitudes are at most
of order $p^2$.
At low energies, the interaction among the
Goldstone bosons thus becomes weak --- pions of zero energy do not interact at
all. This is in marked
contrast to the interaction among the quarks and gluons which is strong at low
energies, because QCD is asymptotically free.
The qualitative difference is crucial for chiral perturbation theory to be
coherent: in this framework, the interaction among the
Goldstone bosons is treated as a perturbation. The opposite behaviour in the
underlying theory prevents a perturbative low energy analysis.

\section{Effective
Lagrangian}\label{lag}\setcounter{equation}{0}

The reformulation of the current algebra technique in the language of an
effective
field theory is discussed in detail in the literature
\cite{pioneer,Callan,Pagels,Weinberg1979}. The
translation exclusively involves general, purely kinematical considerations and
does not leave anything to be desired. I review this only very briefly, to set
up notation.

The one-particle-reducible contributions, which describe the pole
terms occurring in the various Green functions, may be viewed as tree graphs of
a field theory, with pion fields as basic variables.
Since the Goldstone bosons do not carry spin, they are described by
scalar fields, which I denote by $\pi^a(x)$:
the fields are in one-to-one correspondence with the massless
one-particle-states $\mid\!\pi^a(p)\!>$ occurring in the spectrum of asymptotic
states.

In this language,
lines connecting different vertices represent Feynman propagators of the
pion field,
\be
\label{eff1}
<\!0\!\mid\! T\{\pi^a(x) \pi^b (y)\}\!\mid\! 0\!> =
\frac{1}{i}\delta^{ab}\Delta_0(x-y)\;\;\;,\:\;\;
\Delta_0(z)=\!\int\!\!\frac{d^d\!p}{(2\pi)^d}\frac{e^{-ipz}}{-p^2-i\epsilon}
\ee
Vertices which exclusively join pion
lines represent interaction terms occurring in the Lagrangian of the effective
pion field theory, i.e., in the effective Lagrangian. It is important here
that
the vertices admit a Taylor series expansion in powers of the momenta.
In the language of the effective field theory, the
momenta correspond to derivatives of the fields --- the various
terms occurring in the Taylor series represent local interaction terms,
containing the pion fields and their derivatives. A
momentum independent vertex joining four pion lines, e.g., corresponds to an
interaction term of the form $g_{abcd}\pi^a\pi^b\pi^c\pi^d$, while a vertex
involving two powers of momenta is represented by an interaction of the
type $g^{\;\prime}_{abcd}\partial_\mu \pi^a\partial^\mu
\pi^b\pi^c\pi^d$. The translation of the various vertices into
corresponding terms of the effective Lagrangian is trivial: if the vertex
in question joins $P$ pion lines and involves a polynomial in the momenta of
degree $D$, the corresponding term in the effective Lagrangian contains $P$
pion fields and, altogether, $D$ derivatives. Including the standard kinetic
term, which characterizes the propagator (\ref{eff1}), the Lagrangian takes the
form\be
\label{eff6r}
{\cal L}_{\eff}=\mbox{$\frac{1}{2}$}\partial_\mu\pi^a\partial^\mu\pi^a +
v^0(\pi) + v^1_{ab}(\pi)\partial_\mu\pi^a\partial^\mu\pi^b+\ldots
\ee
The Taylor series $v^0(\pi)= \mbox{$\frac{1}{2}$}M^2\pi^a\pi^a
+\mbox{$\frac{1}{3!}$}\,g_{abc}\pi^a\pi^b\pi^c
+\mbox{$\frac{1}{4!}$}\,g_{abcd}\pi^a\pi^b\pi^c\pi^d +\ldots $ yields all
vertices which are momentum independent. The symmetry, of course,
forbids a pion mass term, $M=0$. In fact, as discussed in the preceding
section,
current conservation implies that all of the vertices vanish at zero momentum.
Hence, the effective Lagrangian does not contain any interaction terms
without derivatives, $v^0(\pi)=0$ --- the leading terms in the low energy
expansion of the various vertices
are of $O(p^2)$. The function $v^1_{ab}(\pi)$ collects all of these. The Taylor
expansion of $v^1_{ab}(\pi)$ starts with a term quadratic in $\pi$; the
corresponding Taylor coefficient determines the constants $c^1_{a_1\ldots a_4}$
and $c^2_{a_1\ldots a_4}$, which, as discussed in the preceding section,
account
for the leading terms in the low energy expansion of the four-pion-vertex, etc.
The effective Lagrangian simply collects the information about the various
vertices --- no more, no less.

The coupling of the pions to the currents may also be
accounted for in the effective Lagrangian. The vertex which links the
current to a single pion, e.g., is described by the term
$-F^a_if_\mu^i\partial^\mu\pi^a$, which is linear, both, in the external
fields $f_\mu^i(x)$ and in $\pi^a(x)$. Vertices involving several pion
legs or several currents correspond to terms in the effective Lagrangian which
contain a corresponding number of pion or external fields. The full
effective Lagrangian, which collects the purely pionic vertices as well as
those which describe the interactions with the external fields, is of the form
\be
\label{eff5}
{\cal L}_{\eff}={\cal L}_{\eff}(\pi ,\partial \pi ,\partial^2\pi
,\ldots;f,\partial f,\ldots )\fs
\ee
The general vertex occurring in
this Lagrangian is of the type $\partial^D\!f^E\pi^P$,
where $D$ is the total number of derivatives, $E$ specifies the number of
external fields and
$P$ counts the pion fields entering the interaction term in question. It is
convenient to define the order of the vertex as $O=D+E$, i.e., to treat the
external fields as small quantities of the same order as the momentum,
$f\propto\partial\propto p$. The Lagrangian then consists of a series of
terms\footnote{If the dimension is even, Lorentz invariance only permits
terms of even order.}
with $O=2, 3, \ldots\,$,\be
\label{eff5a}
{\cal L}_{\eff}={\cal L}_{\eff}^{(2)}+{\cal L}_{\eff}^{(3)}+{\cal
L}_{\eff}^{(4)} +\ldots\ee
Note that, in this ordering of the vertices, the number $P$ of pion fields
is left open --- the term ${\cal L}_{\eff}^{(2)}$, e.g., contains vertices
with arbitrarily many pion fields; it collects the purely pionic contributions
to
the effective Lagrangian with two derivatives (the kinetic energy and
the term $v^1_{ab}(\pi)\partial_\mu\pi^a\partial^\mu\pi^b\,$), vertices which
involve one external field and one derivative (such as the term
$-F^a_if_\mu^i\partial^\mu\pi^a\,$), as well as contributions with two
external fields and no derivatives.
The ordering of the various vertices amounts to a generalized
derivative expansion of the effective Lagrangian.

The virtue of the representation in terms of effective fields is that the
Feynman graphs of a local field theory automatically obey the cluster
decomposition property: whenever a given number of pions and currents meet, the
same
vertex occurs, irrespective of the remainder of the diagram. By construction,
the one-particle-reducible contributions to the generating functional are given
by the tree graphs of the effective pion field theo\-ry. Moreover, the
effective theory also provides for a very simple representation of the
multipion exchange contributions required by clustering:
these are described by graphs containing
loops \cite{Pagels} \cite{Weinberg1979}. The sum of all contributions,
involving
the exchange of an arbitrary number of pions between the various vertices is
given by the sum over all Feynman graphs of the effective theory, which is to
be treated in the standard manner, as a {\it quantum} field theory: while
the tree graphs represent the classical limit, graphs with loops describe the
quantum fluctuations.
Accordingly, the representation of the generating functional in terms of
effective fields takes the standard form of a Feynman path
integral
\be\label{eff7} e^{i\,\Gamma\{f\}}=
{\cal Z}^{-1}\!\int
[d\pi]\; e^{ i\!\int\!d^d\!x {\cal L}_{\eff}(\pi ,\partial \pi,\ldots
;f,\partial f,\ldots )}
\ee
where ${\cal Z}$ is the same integral, evaluated at $f=0$.
This formula
represents the link between the underlying and the effective theories: the
quantity $\Gamma\{f\}$ on the left hand side is the generating functional of
the Green functions formed with the current operators of the underlying theory,
while the right hand side exclusively involves the effective field theory. The
pion
pole dominance hypothesis formulated in section \ref{ppd} implies that the
two sides coincide, order by order in the low energy expansion.

As pointed out by Weinberg \cite{Weinberg1979}, the low energy
expansion of the path integral (\ref{eff7}) may be analyzed
perturbatively. To any given, finite order in the momenta, (i) only graphs with
a limited
number of loops contribute and (ii) the derivative
expansion of the effective Lagrangian is needed only to the corresponding
order. More specifically, a graph $\gamma$ with $L$ loops generates
a contribution to the generating functional of order $O(p^{O_\gamma})$, with
$O_\gamma=\sum_{v\in\gamma}(O_v-2) +(d-2)L$. The sum extends over all vertices
of the graph and $O_v$ is the order of the vertex $v$. The leading contribution
stems from the tree graphs of ${\cal L}_{\eff}^{(2)}$, i.e., from
graphs which exclusively involve vertices
with $O_v=2$ and do not contain loops, $L=0$. The current algebra
calculations performed in the 1960's concern this leading order of
the expansion; in these calculations, only the first term in the derivative
expansion of the effective Lagrangian is needed. At first
nonleading order, graphs containing one loop matter and the next term
occurring in the derivative expansion of the Lagrangian also contributes, etc.
Note that the suppression of the loop graphs depends on the dimension of
space-time: in four dimensions, the one-loop graphs
are smaller by two powers of momentum as compared to the tree graphs, while in
$d=3$, they are suppressed only by one power of momentum. In two
dimensions, loops are not suppressed at all --- it is impossible to analyze
the low energy structure of two-dimensional models in terms of an
effective field theory.

I add a few remarks concerning the properties of the measure $[d\pi]$,
referring to the literature \cite{measure} for a more detailed discussion.
In the language of the path integral, the measure
on the space of field configurations is the essential element in the step from
the classical field theory to the corresponding
quantum field theory. An explicit specification of the measure requires
regularization. In the perturbative domain one is concerned with here, it is
well-known that
any regularization procedure may be used and gives rise to the same result
when the cutoff is removed. The measure, i.e., the integrand of
the path integral,
does, however, depend on the regularization.

In dimensional regularization,
the measure takes a remarkably simple form. If this cutoff
procedure is used,
the path integral may be evaluated in the
standard manner, decomposing the Lagrangian into kinetic term plus
interaction and evaluating the latter perturbatively, according to the
Feynman rules. Equivalently, the dimensionally regularized measure may be
defined through the standard formula for Gaussian integrals,
\bea\label{high11}
\int[d\pi]e^{i\frac{1}{2}\!\int\!d^d\!x
\partial_\mu\pi\partial^\mu\pi}\;\left\{\mbox{$\int$}d^d\!x\,\psi_a(x)\pi^a(
x ) \right\}^{2n}=\hspace{6em}	\no \hspace{6em} =
(2n-1)!!\left\{\mbox{$\int$}d^d\!xd^d\!y\,\psi_a(x)\,\mbox{$\frac{1}{i}$}\,
\delta^{ab} \Delta_0(x-y)\,\psi_b(y)\right\}^n \fs\eea
As the formula holds for an arbitrary set of test functions $\psi^a(x)$, it
fully determines the path integral over the
various contributions arising in the perturbative expansion.
The reasons for
the simplicity of the measure in dimensional regularization are that this
method (i)
preserves the symmetries of the theory and (ii) avoids the occurrence of power
divergences (regularization dependent terms
which grow with a power of the cutoff). For other cutoff procedures,
the expression for the measure involves contributions
proportional to $\delta (0)\sim
\Lambda^d$ or to a derivative thereof. In
dimensional regularization, such terms vanish ab initio,
$\delta(0)=\partial_\mu\delta(0)=\partial_{\mu\nu}\delta(0)=\ldots =0$.

In the present context, the crucial property of the measure is gauge
invariance: if the action functional
$S_{\eff}\{\pi,f\}=\int\!d^d\!x{\cal L}_{\eff}[\pi,f]$ is invariant under
a simultaneous gauge transformation of the fields $\pi,f$, then the
corresponding path integral is a gauge invariant functional of the external
fields. In contrast to the chiral symmetries of fermionic theories, which
only hold at the classical level and do not represent symmetries of the
measure, the quantum fluctuations of the effective fields do maintain gauge
invariance.

\section{Invariance theorem}
\label{it}\setcounter{equation}{0}

Chiral perturbation theory is based on the
{\em assumption} that the effective Lagrangian is invariant under a
simultaneous
gauge transformation of the fields $f_\mu^i(x)$ and $\pi^a(x)$. While the
transformation law of the external fields is specified in equation
(\ref{found5}), the pion field transforms with a nonlinear representation of
G. One usually replaces the variables $\pi^a(x)$ by a matrix field $U(x)$ with
linear
transformation properties. In the context of QCD, e.g., one may work with
the unitary matrix $U=\exp(i\pi^a\lambda_a/F_\pi)$, for which the
transformation law reads $U\stackrel{g}{\rightarrow}V_{\mbox{\scriptsize
R}}UV^\dagger_{\mbox{\scriptsize L}}$. Invariance of
the effective Lagrangian ${\cal L}_{\eff}(U,\partial
U,\ldots;f,\partial f,\ldots)$ under a simultaneous gauge
transformation of the fields $U(x)$ and $f_\mu^i(x)$
is sufficient to insure a gauge invariant path
integral, but is it necessary? In the following it is shown
that this question can be answered
affirmatively: For Lorentz invariant theories in four dimensions,
the effective Lagrangian is gauge
invariant to all orders of the derivative expansion. I refer
to this assertion as an invariance theorem. The proof makes
essential use of Lorentz invariance. In nonrelativistic theories, the time
components of the currents may develop an expectation value. If this happens,
the corresponding effective Lagrangian is gauge invariant only up to a total
derivative \cite{Ferromagnet}.
Also, for Lorentz invariant theories in three dimensions, the assertion
requires a slight modification, related to the occurrence of Chern-Simons
terms.

For the following general discussion, it is more convenient not to work with a
matrix field, but to view the pion field variables as
coordinates of the quotient space
G/H \cite{Callan}. The elements of this space are the equivalence classes of
the group G under rightmultiplication with the subgroup H: the
elements $g_1,g_2\in \mbox{G}$ belong to the same class if $g_1^{-1}g_2\in
\mbox{H}$. Picking a representative element $n\in\mbox{G}$ in each one of the
equivalence classes, every element of the group may uniquely be decomposed as
$g=nh$, with $h\in\mbox{H}$. The group acts on G/H through
leftmultiplication: the image of the class belonging to $n$ is the equivalence
class of $gn$. The corresponding representative element $n^\prime$ is obtained
from the decomposition $gn=n^\prime h$.

The space G/H is
of dimension $d_{\mbox{\scriptsize G}}
-d_{\mbox{\scriptsize H}}=\mbox{N}_{\mbox{\scriptsize GB}}$. One thus needs as
many coordinates to label the elements of G/H as there are Goldstone bosons.
Identifying the
variables $\pi^a$ with the coordinates on G/H, the pion field $\pi^a(x)$ may be
viewed
as a mapping from Minkowski space into G/H. The representative elements are in
one-to-one correspondence with the field variables, $n=n_\pi$. The action of
the group on
G/H thus induces a map in the space of the field variables:
\be\label{it1}
\pi\stackrel{g}{\rightarrow}\varphi(g,\pi)\fs\ee
I refer to this map as the {\em canonical} transformation law of the pion
field. In terms of the
corresponding representative elements, the canonical map is defined by
$gn_\pi=n_{\varphi(g,\pi)}h$. The canonical transformation law is equivalent to
the one mentioned above, involving a matrix representation of the pion field,
and readily extends to gauge
transformations; it suffices to allow the group element which enters the
transformation to depend on $x$:
$\pi(x)\stackrel{g(x)}{\rightarrow}\varphi\left(g(x),\pi(x)\right)$.

The effective Lagrangian is a
function of the fields $\pi^a(x),f_\mu^i(x)$ and their derivatives at one and
the same point of space-time, ${\cal L}_{\eff}=
{\cal L}_{\eff}(\pi,\partial\pi,\ldots;f,\partial f\ldots)$. In the following,
local
functions of this type repeatedly occur. I simplify the notation, using square
brackets to indicate arguments which also enter through their derivatives. In
this notation, the Lagrangian is written as ${\cal L}_{\eff}[\pi,f]$.
Note the
difference between these {\em local functions} and nonlocal {\em
functionals} such as the
classical action of the effective field theory,\be\label{inv3}
S_{\eff}\{\pi,f\}\equiv\int\!d^d\!x{\cal L}_{\eff}[\pi,f]\co\ee
which depends on the values of the fields $\pi,f$
throughout space-time. I use curly brackets for the arguments of
such functionals.

The invariance theorem is based on the following assertions, which will
be established one after the other:

{\bf A}. There exists a mapping of the pion
field,
$\pi\stackrel{g}{\rightarrow}\phi[g,\pi,f]$, such that, together with the
standard gauge transformation of the external fields,
the action functional remains invariant,\be\label{inv7}
S_{\eff}\{\phi[g,\pi,f],T(g)f\}=S_{\eff}\{\pi,f\}\fs\ee
The function $\phi[g,\pi,f]$ is of the same structure as the effective
Lagrangian: a sequence of local terms involving an increasing number of
derivatives of the fields $g,\pi,f$ at the same point of space-time.

{\bf B}. The map $\phi[g,\pi,f]$
represents a nonlinear realization
of the group G, i.e., obeys the composition law\be\label{inv7a}
\phi[g_2g_1,\pi,f]=\phi[g_2,\phi[g_1,\pi,f],T(g_1)f]\fs\ee

{\bf C}. With a suitable change $\pi^a
\rightarrow \psi^a[\pi,f]$ of the field variables,
the map may be brought to the canonical form
specified above.
In these coordinates, the transformation law of the pion field is fully
determined by the
geometry of the groups G and H and is independent of the interaction.
Gauge invariance then takes the
form\be\label{inv7d}
S_{\eff}\{\varphi(g,\pi),T(g)f\}=S_{\eff}\{\pi,f\}\fs\ee

{\bf D}. In four dimensions, the
effective Lagrangian itself is gauge invariant,
\be\label{inv7b}
{\cal L}_{\eff}[\varphi(g,\pi),T(g)f]={\cal L}_{\eff}[\pi,f]\fs\ee
In three dimensions, this is true only up to the possible occurrence of a
Cherns-Simons term of order $O(p^3)$,
\bea\label{inv7c}
{\cal L}_{\eff}[\pi,f]&\hspace{-0.5em}=&\hspace{-0.5em}
{\bar{\cal L}}_{\eff}[\pi,f]
+{\cal L}_{\mbox{\scriptsize CS}}[f]\no {\cal L}_{\mbox{\scriptsize CS}}[f]
&\hspace{-0.5em}=&\hspace{-0.5em}
c\epsilon^{\lambda\mu\nu}
\mbox{tr}\{f_\lambda\partial_\mu f_\nu
-\mbox{$\frac{2}{3}$}if_\lambda f_\mu f_\nu\}\co\eea
where $\bar{\cal L}_{\eff}[\pi,f]$ is gauge invariant. The Chern-Simons
term only involves the external fields. The integral $\int\!d^3\!x{\cal
L}_{\mbox{\scriptsize CS}}[f]$ is gauge invariant, but the integrand is not.

In the next two sections, these assertions are shown to hold true at the
leading order of the low energy
expansion. The extension of the proof to all orders is discussed in section
\ref{high}.

\section{Leading order}
\label{tree}\setcounter{equation}{0}
Consider first the leading order of the low energy expansion.
As discussed in section \ref{lag}, the expansion starts with the tree graph
contributions generated by ${\cal L}_{\eff}^{(2)}$. The general Lorentz
invariant expression for this part of the Lagrangian is of the form
\be\label{eff16}
{\cal L}_{\eff}^{(2)} =  \mbox{$\frac{1}{2}$}g_{ab}(\pi )\,
\partial_\mu \pi^a \partial^\mu
\pi^b - h_{ai}(\pi )f^i_\mu\partial^\mu \pi^a +\mbox{$\frac{1}{2}$}k_{ik}(\pi
)f^i_\mu f^{k\,\mu}
+\,l_a(\pi)\, \raisebox{0.75mm}{\fbox{\rule[0.5mm]{0mm}{0mm}\,}}\,\pi^a +
m^i(\pi)\,\partial^\mu f^i_\mu \fs
\ee
The conservation of energy and momentum implies that total
derivatives
do not contribute. Hence one may integrate the last two terms by parts and
absorb them in the first two: without loss of generality, one may set
$l_a(\pi)=m^i(\pi)=0$.

The tree graphs
describe the theory in the classical limit. More precisely, the sum of all tree
graph
contributions to the path integral (\ref{eff7}) is given by the classical
action, evaluated at the extremum,
\be
\label{eff20}
\Gamma\{f\}=\mbox{extremum}\hspace{-2.5em}
\raisebox{-1em}{$\pi $}\hspace{2em}\;S_{\eff}^{(2)}\{\pi,f\}+O(p^3)\ee
Accordingly, the issue boils down to a problem of classical field theory:
what are
the conditions to be satisfied by the Lagrangian, in order for the classical
action, evaluated at the extremum, to be gauge invariant?

At the extremum, the pion field obeys the classical equation of
motion,
\be\label{tree1}
\frac{\delta S_{\eff}^{(2)}\{\pi,f\}}{\delta \pi^a(x)}
=\frac{\partial {\cal L}^{(2)}_{\eff}}{\partial \pi^a}
 -\partial_\mu \!\left(\frac{\partial {\cal L}^{(2)}_{\eff}}{\partial(
\partial_\mu \pi^a)} \right)
=0\fs\ee
As the value of the action at the extremum is stable against variations of
the pion field, the change in the classical solution generated by an
infinitesimal gauge transformation of the external fields does not contribute.
Gauge invariance thus requires
\be \label{eff23}
D_\mu\frac{\delta S^{(2)}_{\eff}\{\pi,f\}}{\delta f^i_\mu (x)}=
D_\mu \!\left(\frac{\partial {\cal L}^{(2)}_{\eff}}{\partial f^i_\mu }\right)
 =0 \fs
\ee
Hence, the pion field simultaneously obeys two differential equations,
\bea \label{gu2}
g_{ab}\raisebox{0.75mm}{\fbox{\rule[0.5mm]{0mm}{0mm}\,}}\,\pi^b
&\hspace{-0.5em}+&\hspace{-0.5em}(\partial_c g_{ab}-
\mbox{$\frac{1}{2}$}\partial_a
g_{bc})\partial_\mu\pi^b \partial^\mu \pi^c +
(\partial_ah_{bi}-\partial_bh_{ai})f_\mu^i\partial^\mu\pi^b
\no&\hspace{-0.5em}  -&\hspace{-0.5em} h_{ai}\partial^\mu
f_\mu^i -\mbox{$\frac{1}{2}$} \partial_ak_{ik}f_\mu^if^{k\mu} =0\\
\label{gu2a}h_{ai}\raisebox{0.75mm}{\fbox{\rule[0.5mm]{0mm}{0mm}\,}}\,\pi^a
&\hspace{-0.5em}+&\hspace{-0.5em}\partial_ah_{bi}\partial_\mu
\pi^a\partial^\mu\pi^b
-\partial_ak_{ik}f^i_\mu \partial^\mu \pi^a
 -k_{ik}\partial^\mu f^k_\mu  \no&\hspace{-0.5em} +&\hspace{-0.5em}
f^l_{\;ik}f^k_\mu
(h_{al}\partial^\mu\pi^a -k_{lm}f^{m\mu})=0\fs \eea
The partial derivatives of the functions $g_{ab}(\pi), h_{ai}(\pi),
k^{ik}(\pi)$ occuring here represent derivatives
with respect to the pion variables, $\partial_a =\partial/\partial\pi^a$.

In analyzing these relations, it is useful to interpret the matrix
$g_{ab}(\pi)$ as a metric
on the manifold G/H. Since the expansion in powers of $\pi^a$ starts with the
contribution from the kinetic term, $g_{ab}(\pi)=\delta_{ab}+\dots\,$, the
metric possesses an inverse $g^{ab}(\pi)$, at least in the vicinity of the
origin. I make use of
the standard bookkeeping, converting covariant indices into contravariant ones
and vice versa by means of the metric (e.g., $h^a_{\;i}=g^{ab}h_{bi}$),
and also make use of the affine connection induced by the metric,
\be \label{gu1c}
\Gamma^{\,c}_{ab}=\mbox{$\frac{1}{2}$} g^{cd}
(\partial_a g_{bd}+\partial_b g_{ad}-\partial_d g_{ab})\fs
\ee

Eliminating $\raisebox{0.75mm}{\fbox{\rule[0.5mm]{0mm}{0mm}\,}}\,\pi^a$
between the two relations (\ref{gu2}) and (\ref{gu2a}), one obtains a
constraint
on the pion field and the first derivatives thereof. At a given point
$x$, these quantities
are, however, independent from one another.
(At a
fixed time, the field $\pi^a$ and its first time derivative,
$\dot{\pi}^a$, represent initial values, which determine the
solution of the equation of motion and are not constrained by it. In
the present context, where the classical solution of interest is the one
selected
by Feynman boundary conditions at $x^0\rightarrow\pm\infty$, these functions
depend on the behaviour of the external
field in the past and in the future, while the constraint only involves the
external field and its first derivatives at the point under consideration.)
Hence the constraint is
consistent with the equation of motion only if the coefficients occurring
therein vanish identically.
This requires
\bea \label{gu6}
&(a)&\;\;d_ih^a_{\;k}-d_kh^a_{\;i}= f^l_{\;ik}h^a_{\;l}\no
&(b)&\;\;\nabla_{\!a}h_{bi}+\nabla_{\!b}h_{ai}=0\hspace{6em}\no
&(c)&\;\;k_{ik}=g^{ab}h_{ai}h_{bk}\co
\eea
where the differential operators
$d_i$ stand for
\be \label{gu5}
d_i =h^a_{\;i}(\pi) \partial_a
\ee
and $\nabla_{\!a}$ is the covariant derivative with respect to the metric
\be \label{gu7}
\nabla_{\!a}h_{bi}= \partial_a h_{bi} -\Gamma^{\,c}_{ab}h_{ci}\fs
\ee
Relation (c)
implies that the functions $k_{ik}(\pi)$ are determined by $g_{ab}(\pi)$ and
$h^a_{\;i}(\pi)$.
The effective Lagrangian may thus be written in the form
\bea\label{gu8}
{\cal
L}^{(2)}_{\eff}&\hspace{-0.5em}=&\hspace{-0.5em}\mbox{$\frac{1}{2}$}g_{ab}(\pi)
D_\mu \pi^aD^\mu\pi^b \\ \label{gu8r}
D_\mu\pi^a&\hspace{-0.5em}\equiv&\hspace{-0.5em}\partial_\mu \pi^a
-h^a_{\;i}(\pi)\,f^i_\mu\fs
\eea
The purely pionic vertices are
described by the metric $g_{ab}(\pi)$; the coupling to
the external field in addition involves the functions $h^a_{\;i}(\pi)$. In
particular, the vacuum-to-pion matrix elements of the currents are given by
$F^a_i=h_{ai}(0)$. The tree graphs
generated by ${\cal L}^{(2)}_{\eff}$ satisfy the Ward identities if and only if
$g_{ab}(\pi)$ and $h^a_{\;i}(\pi)$ obey the first order differential equations
$(a)$ and $(b)$.

These relations insure gauge invariance of the Lagrangian
${\cal L}^{(2)}_{\eff}$. Indeed,
consider the infinitesimal gauge transformation of the
external field specified in (\ref{found6})
and subject the pion field to the change
\be \label{gu8b}
\delta\pi^a(x)=g^i(x)\,h^a_{\;i}\left(\pi(x)\right)\fs
\ee
The relation $(a)$ then implies
that the quantity $D_\mu\pi^a$ defined in (\ref{gu8r}) transforms covariantly,
\be \label{gu8c}
\delta\left\{D_\mu\pi^a\right\}=(g^i\,\partial_{\,b}h^a_{\;i})\,D_\mu\pi^b\fs
\ee
The deformation in the metric is given by $\delta
g_{ab}=\partial_cg_{ab}\delta\pi^c$. Collecting terms, the change in the
effective Lagrangian may be expressed in terms of the covariant derivative
defined in (\ref{gu7}),
\be \label{gu8d}
\delta {\cal L}_{\eff}^{(2)} = g^i\,\nabla_{\!a} h_{bi}\,
D_\mu\pi^aD^\mu\pi^b\fs \ee
Finally, since only the symmetric part of the derivative contributes, the
relation $(b)$ entails the invariance property claimed above,
$\delta {\cal L}_{\eff}^{(2)} = 0$.

This verifies the invariance theorem at leading order of the low energy
expansion, except for the claim that the transformation law of the pion field
takes the canonical form specified in section \ref{it}.

\section{Differential geometry of the Goldstone bosons}
\label{geo}\setcounter{equation}{0}
Condition (b) states that the vectors $h^a_{\;i}(\pi)$ represent Killing
vectors of the differential geometry characterized by the metric $g_{ab}(\pi)$.
This geometry thus admits a group of isometries. Moreover, relation (a) shows
that the structure constants of the isometry group are those of G: the
metric $g_{ab}(\pi)$ describes a symmetric space.

The relation (a) implies that the differential operators $d_i$
obey the
commutation rule $[d_i,d_j]=f^k_{\;ij}d_k$, i.e., the operators $id_i$ form a
representation of the Lie algebra of G.
Any representation of the Lie algebra may be
integrated to a representation of the group, at least in a finite
neighbourhood of the unit element. The resulting representation
$O(g)$ of G obeys the composition law $O(g_2)O(g_1)=O(g_2g_1)$,
provided all of the
elements are in the neighbourhood of unity. If the group is multiply connected,
there are inequivalent paths connecting the unit element with $g$,
such that the composition law may fail to hold globally. In the context of the
low energy expansion, the global properties are, however, not relevant.
The evaluation of the path integral to any given order of the low energy
expansion only involves vertices with a limited number of pion fields. These
vertices are the
Taylor coefficients of the functions $g_{ab}(\pi), h^a_{\;i}(\pi),
k^{ik}(\pi)$. The entire analysis thus only concerns the
vicinity of the unit element, where the composition law holds as it stands.
I refrain from repeatedly mentioning this proviso and simply speak of the
group when referring to elements contained in a finite neighbourhood of
unity.

Since $O(g)$ is a representation of the group, the function
$\bar{\varphi}^a(g,\pi)\equiv O(g^{-1})\pi^a$ obeys the composition rule
\be\label{tree2}
\bar{\varphi}(g_2,\bar{\varphi}(g_1,\pi))=\bar{\varphi}(g_2g_1,\pi)\ee
Remarkably, this property determines the function $\bar{\varphi}(g,\pi)$
essentially uniquely \cite{Callan}. Consider the image of the origin,
$\bar{\varphi}(g,0)$. The infinitesimal form (\ref{gu8b}) of the map shows that
the origin is invariant under H: the quantity $h_{ai}(0)=F^a_i$ vanishes if
the index $i$ belongs to the subalgebra {\bf H}.
Accordingly, $\bar{\varphi}(h,0)=0,\,\forall \;h\in\mbox{H}$. The
composition law (\ref{tree2}) then implies
$\bar{\varphi}(gh,0)=\bar{\varphi}(g,0)$, i.e., the value of the function
only depends on the equivalence class. Hence, $\bar{\varphi}(g,0)$ maps the
elements of G/H into the
space of pion field variables.
The mapping is unique, except for the freedom in the choice of coordinates
on G/H and in the space of field variables.
Without loss of generality, one may choose variables such that the function
$\bar{\varphi}(g,0)$ coincides with the
canonical map $\varphi(g,0)$ (which, of course, also depends on the
parametrization). The composition law then implies that the two
maps coincide everywhere on G/H, $\bar{\varphi}(g,\pi)=\varphi(g,\pi)$.
This verifies
the claim that the transformation law of the pion field may be brought to
canonical form.

The argument just given shows that the functions $h^a_{\;i}(\pi)$, which
collect the vertices associated with the coupling of the Goldstone bosons
to the currents, are purely
geometrical quantities, determined by the structure of the groups G and H:
these functions represent the infinitesimal form of the map $\varphi(g,\pi)$.

Next, consider the metric. The Killing condition
(b) represents the infinitesimal form of the relation
\be\label{gu50}
\partial_a\varphi^c(g,\pi)\partial_{\,b}\varphi^d(g,\pi)g_{cd}
\left(\varphi(g,\pi) \right) =g_{ab}(\pi) \co
\ee
which states that the line element
$ds^2=g_{ab}(\pi)d\pi^ad\pi^b$ is invariant under the mapping
$\pi\stackrel{g}{\rightarrow}\varphi(g,\pi)$.
Every point in the neighbourhood may be reached from the origin with a
suitable choice of $g$, such that the above
relation fixes
the form of the metic in terms of the matrix $g_{ab}(0)$. The standard
normalization of the pion field, $g_{ab}(0)=\delta_{ab}$, suggests
that the metric is fully determined by group geometry. This impression is
misleading, however, because the freedom in
the choice of variables is in effect exploited twice: first, it was argued that
the form of the Killing vectors $h^a_{\;i}(\pi)$ only depends on the choice of
coordinates on G/H and, now, the same freedom is used to identify the metric at
the origin with the euclidean metric of the tangent space. The scalar
product of the Killing vectors is independent of the
choice of coordinates and is determined by the decay constants,
\be\label{tree4}
g_{ab}(0)h^a_{\;i}(0)h^b_{\;k}(0)=\delta_{ik}F_{(i)}^2\fs\ee
If one exploits the freedom in the choice of coordinates by setting
$g_{ab}(0)=\delta_{ab}$, then the Killing vectors do carry information which
goes beyond group geometry --- the decay constants are then given by the
components of the Killing vectors at the origin.

The main point here is that, up to parametrization, the leading term in the
derivative expansion of the effective Lagrangian is fully determined by the
decay constants, which play the role of effective coupling
constants.
The number of independent effective couplings is determined by the
transformation properties of the Goldstone bosons under H: every
irreducible multiplet requires its own decay constant \cite{Boulware}.
The parametrization used
is irrelevant --- the generating functional is given by the extremum of the
classical action, which is invariant under a change of variables.

I briefly comment on the geometric significance of the result.
The {\it intrinsic geo\-metry} of a
compact
Lie group is invariant under both, right- and lefttranslations. For simple
groups, this geometry is fixed up to an overall normalization
constant, which may be chosen such that the inner product of the Killing
vectors agrees with the Cartan metric of the Lie algebra. By
projection, the geometry of the group also induces an intrinsic metric on
the quotient space G/H, which I denote
by $\bar{g}_{ab}(\pi)$.

If the pions transform irreducibly under
H, the metric occurring in the effective Lagrangian indeed coincides with the
intrinsic geometry of G/H, except for an overall factor:
$g_{ab}(\pi)=F^2\bar{g}_{ab}(\pi)$.
In the general case, however, the induced
metric is
not the one which matters. For the extreme situation
of a totally broken symmetry, e.g., where H only contains the unit
element, the quotient space G/H is the group itself and the
induced metric coincides with the intrinsic geometry of the group. In
that case, the metric of the
effective Lagrangian, however, involves as many independent decay constants as
there are pions, indicating that the
relevant geometry is
less symmetric than the intrinsic one. In the general case,
the geometry relevant for the Goldstone bosons
is the one induced on the
quotient space G/H by the general metric on the group which is
{\it leftinvariant under} G, {\it but rightinvariant only under} H.
The metric relevant for the Lagrangian
is obtained by decomposing the intrinsic geometry of G/H into a sum
of
contributions $\bar{g}_{ab}(\pi)=\sum_i g^{(i)}_{ab}(\pi)$ (at the origin, the
decomposition corresponds to the various orthogonal subspaces of {\bf K},
which transform irreducibly under H).
The decay
constants stretch the different components of the intrinsic line element
by different factors, replacing the above sum by $
g_{ab}(\pi)=\sum_i F_{(i)}^2g^{(i)}_{ab}(\pi)$.
The geometry of the manifold G/H thus resembles an ellipsoid, the
decay constants
playing the role of the semi-axes. The analogy is not perfect, however:
the metric $g_{ab}(\pi)$ still possesses G as a group
of isometries, which acts transitively on the manifold, such that the geometry
in the vicinity of any given point is the same as around
the origin.

\section{Higher orders}\label{high}
\setcounter{equation}{0}

In the present section,
the above analysis of the leading term ${\cal
L}^{(2)}_{\eff}$ is extended to all orders of the derivative expansion, using
induction.
The induction hypothesis is
that the invariance
theorem holds up to and including ${\cal L}^{(n)}_{\eff}$.
For the low energy representation of the generating functional
to order $p^{n+1}$, the action entering the path integral
(\ref{eff7}) may be truncated at
\bea\label{ho9}
S_{\eff}\{\pi,f\}_{n+1}&\hspace{-0.5em}=&\hspace{-0.5em}
S^{(2)}_{\eff}\{\pi,f\}+\ldots+S^{(n+1)}_{\eff}\{\pi,f\}\no
S^{(m)}_{\eff}\{\pi,f\}&\hspace{-0.5em}\equiv&\hspace{-0.5em}
\int\!d^d\!x {\cal L}^{(m)}_{\eff}[\pi,f]\fs\eea
Moreover, the term $S^{(n+1)}_{\eff}\{\pi,f\}$ exclusively enters through tree
graph contributions. Loop graphs only involve vertices from those
parts of the action, which, by the induction hypothesis, are gauge
invariant. Hence the path
integral is gauge invariant to order $p^{n+1}$ if and only if the
tree graphs are. Accordingly, the issue again reduces to a problem of classical
field theory: determine the general solution of the simultaneous differential
equations
\be\label{ho10}
\frac{\delta S_{\eff}\{\pi,f\}_{n+1}}{\delta\pi^a(x)}=0\;\;\;\;,\;\;\;\;
D_\mu\frac{\delta S_{\eff}\{\pi,f\}_{n+1}}{\delta f^i_\mu (x)}=0
\fs\ee
The core of the proof consists of an analysis of these two
equations, which proceeds along the list of assertions made
in section \ref{it}. I briefly outline the essence of the argument, referring
to the appendix for the details.

{\bf A}. The first step is the construction of the
map $\phi[g,\pi,f]$ (appendix \ref{map}). The construction merely extends the
discussion given
in section \ref{tree}: for the two differential equations to be consistent with
one another, they must be linearly dependent. Roughly speaking, the function
$\phi[g,\pi,f]$ is the coefficient occurring in the relation which expresses
this linear dependence.
As compared to the situation encountered in the preceding section,
the only complication brought about by the occurrence of
higher derivatives is that the transformation law of the pion field is
modified and now involves derivatives of the fields $g^i(x),\pi^a(x)$, as well
as the external fields $f_\mu^i(x)$.

{\bf B}. The next step concerns the assertion that the mapping
$\pi\stackrel{g}{\rightarrow}\phi[g,\pi,f]$
yields a representation of the group.
Denote the solution of the equation of motion by $\pi_f(x)$. The invariance
of the action implies that the transformed solution, $\phi[g,\pi_f,f]$ obeys
the
equation of motion belonging to the transformed external fields, $T(g)f$. Since
the solution is unique,
this implies $\phi[g,\pi_f,f]=\pi_{T(g)f}$. Now, the transformation law of
the external fields does satisfy the composition law,
$T(g_2)T(g_1)f=T(g_2g_1)f$. Hence
$\phi[g_2,\phi[g_1,\pi_f,f],T(g_1)f]
= \phi[g_2g_1,\pi_f,f]$ --- on the solution of the equation of motion, the
composition rule is valid. The argument is extended to arbitrary
configurations of the pion field in appendix \ref{cl}.

{\bf C}. The proof of the third assertion is more involved. It exploits the
fact
that the composition law strongly constrains the form of the local function
$\phi[g,\pi,f]$. The general solution of this constraint
shows that the map differs from the canonical one only by a change of
variables (appendix \ref{can}).
This then completes the inductive argument, demonstrating that the assertions
{\bf A,B,C} hold to all orders: the functional $S_{\eff}\{\pi,f\}$ is
invariant under the canonical transformation of the fields $\pi$ and
$f$.

{\bf D}. The consequences for the effective Lagrangian may then be derived as
follows. Since the element $g=n_\pi^{-1}$ takes the pion field into
the origin, the invariance property (\ref{inv7d}) implies that the action
functional may be expressed in terms of its values at zero field,
\be\label{inv40}
S_{\eff}\{\pi,f\}=S_{\eff}\{0,f_\pi\}\;\;\;,\;\;\;f_\pi=T(n_\pi^{-1})f\fs\ee
The relation shows that the difference
${\cal L}_{\eff}[\pi,f]-{\cal L}_{\eff}[0,f_\pi]$ is a total derivative,
$\partial_\mu \omega^\mu[\pi,f]$, which disappears if the pion field is turned
off, $\partial_\mu \omega^\mu[0,f]=0$. Since one may add
a total derivative to the Lagrangian without changing the content of
the theory, it is legitimate to replace ${\cal L}_{\eff}[\pi,f]$ by
${\cal L}_{\eff}[\pi,f]-\partial_\mu \omega^\mu[\pi,f]$, such
that
\be\label{inv40a}
{\cal L}_{\eff}[\pi,f]={\cal L}_{\eff}[0,f_\pi]\fs\ee

The configuration $\pi=0$
is invariant under the subgroup H; gauge invariance thus requires
\be\label{inv41}
S_{\eff}\{0,T(h)f\}=S_{\eff}\{0,f\}\;\;\;\;\;\forall \,h\in\mbox{H}\fs\ee
Conversely, this property insures that the corresponding full action,
specified in equation (\ref{inv40}), is gauge invariant under the full group.
So, what remains to
be done is to analyze the implications of gauge invariance with respect to
H at zero pion field.

Under the action of the subgroup, the external vector field $f_\mu$
associated
with the currents of G does not transform irreducibly. Decompose the field
according to $f_\mu=v_\mu+a_\mu$, where
the first part contains those components
which belong to the subspace ${\bf H}$ of the Lie algebra,
$v_\mu=\sum_{i\in {\bf H}}t_i\,f^i_\mu $, while
$a_\mu =\sum_{i\in {\bf K}}t_i\,f^i_\mu $
represents the remainder (in QCD, $v_\mu$ and $a_\mu$ are the
vector and axial vector fields, respectively).
The field $v_\mu $ transforms like a gauge field of H,
while $a_\mu$ transforms homogeneously, according to the
same representation as the pions,
\be\label{high5}
T(h)v_\mu =D(h)v_\mu D(h)^{-1}-i\partial_\mu D(h)D(h)^{-1}\;\;\;,\;\;\;
T(h)a_\mu=D(h)a_\mu D(h)^{-1}\fs\ee

Consider first the dependence of the Lagrangian on $a_\mu$.
The relation (\ref{inv41}) implies that the
variational derivative
\be\label{inv42}
A^\mu[v,a]
=\frac{\delta S_{\eff}\{0,v,a\}}
{\delta a_\mu (x)}\ee
transforms according to
\be\label{inv43}
A^\mu
[T(h)v,T(h)a]
=D(h)A^\mu[v,a]D(h)^{-1}\fs\ee
The difference between the full action and the one for $a_\mu=0$ is given by
the integral over the derivative $\frac{d}{dt}S_{\eff}\{0,v,ta\}$ from
$t=0$ to $t=1$,
\be\label{inv44}
S_{\eff}\{0,v,a\}=
S_{\eff}\{0,v,0\}+\int\!\!d^d\!x\!\int_0^1\!\!\!dt\;
\mbox{tr}(a_\mu A^\mu [v,ta])
\fs\ee
The action determines the Lagrangian only up to a total
derivative. One may exploit this freedom and set
\be\label{inv44a} {\cal L}_{\eff}[0,v,a]
={\cal L}_{\eff}[0,v,0]
+\int_0^1\!\!\!dt\;\mbox{tr}(a_\mu A^\mu [v,ta])  \fs\ee
The virtue of this choice is that, by construction,
the part of the Lagrangian which involves the field $a_\mu$ is gauge
invariant.

This reduces the matter to an elementary problem of gauge field theory: the
remainder of the action, $S_{\eff}\{0,v,0\}$, exclusively
involves a gauge field, $v_\mu(x) $, with gauge group H. The action is gauge
invariant. What are the implications for the Lagrangian?

The example of the Chern-Simons Lagrangian in $d=3$
shows that gauge invariance of the action does not in general
imply gauge invariance of the Lagrangian. In appendix \ref{gau}, it is shown
that this example is \underline{the} exception: the
effective Lagrangian is gauge invariant, up to a
Chern-Simons term, which may only occur for $d=3$. The proof relies on a
general property
of local differential forms, which is established in appendix \ref{forms}.

Note that gauge invariance
does not fix the form of
the effective Lagrangian completely: gauge invariant total
derivatives may be added and there are point transformations which
preserve the canonical transformation law of the pion field.
Only the leading
term of the derivative
expansion is fully determined by the geometry of the groups G and H.
The remaining freedom in the choice of the field variables is
equivalent to the well-known fact that one is free to modify the higher order
terms by adding gauge invariant multiples of the equation of motion.

\section{Anomalies}
\label{anom}\setcounter{equation}{0}
The generating functional is gauge
invariant only if the Ward identities do not contain anomalies.
I briefly discuss the modification of the preceding
analysis required by the occurrence of
anomalies. For definiteness, I use the nomenclature of QCD.

If there are $\mbox{N}_f$
massless quark flavours, the Hamiltonian is invariant
under the group $\mbox{G}=\mbox{SU}(\mbox{N}_f)_{\mbox{\scriptsize
R}} \times \mbox{SU}(\mbox{N}_f)_{\mbox{\scriptsize L}} $ of global chiral
rotations,
\be\label{found7}
T(g)q(x)_{\mbox{\scriptsize R}}=V_{\mbox{\scriptsize R}}\,
q(x)_{\mbox{\scriptsize R}}\;\;\;\;,\;\;\;\;T(g)q(x)_{\mbox{\scriptsize L}}=
V_{\mbox{\scriptsize L}}\,q(x)_{\mbox{\scriptsize L}}\ee
and the corresponding currents $J^\mu_{i\;{\mbox{\scriptsize R}}}=
\bar{q}_{\mbox{\scriptsize R}}\gamma_\mu \mbox{$\frac{1}{2}$}\lambda_i
q_{\mbox{\scriptsize R}}$, $J^\mu_{i\;{\mbox{\scriptsize L}}}=
\bar{q}_{\mbox{\scriptsize L}}\gamma_\mu \mbox{$\frac{1}{2}$}\lambda_i
q_{\mbox{\scriptsize L}}$ are strictly conserved.\footnote{Note that the
discussion does not include the singlet currents ---
the axial U(1)-current fails to
be conserved, also on account of an anomaly. The effective Lagrangian
analysis may be extended to the Green functions of these currents by
treating the vacuum angle as an external field
\cite{GLNP}.} The generating
functional of massless QCD thus involves two sets of external fields,
$f_\mu=(f_\mu(x)_{\mbox{\scriptsize R}}, f_\mu(x)_{\mbox{\scriptsize
L}})$. In view of the anomalous terms occurring in the
Ward identities for the Green functions formed with the currents, the
generating functional, however, fails to be gauge invariant.
Under an infinitesimal chiral
rotation,
\be\label{found8}
V_{\mbox{\scriptsize R}}={\bf 1}+i\alpha(x)+i\beta(x)\;\;\;\;\;\;
V_{\mbox{\scriptsize L}}={\bf 1}+i\alpha(x)-i\beta(x)\co\ee
the generating functional undergoes the change
\be\label{found9}
\delta\Gamma\{f\}
=-\!\int\!d^4\!x\mbox{tr}\{\beta(x)\,\Omega[f(x)]\}\co\ee
where $\Omega[f]$ is a local
function of $O(p^4)$, formed exclusively with the external fields --- the
explicit expression is not needed here \cite{Bardeen}.

The main point is that
anomalies do not destroy the symmetry of the theory with respect to gauge
transformations
of the external fields --- they merely mo\-di\-fy the transformation law of
the generating functional, replacing the condition $\delta \Gamma\{f\} =0$ by
the constraint (\ref{found9}), which is equally strong.

In the low energy expansion, anomalies only start showing up at first
nonleading order
--- the differential geometry of ${\cal L}^{(2)}_{\eff}$, discussed in
section \ref{geo}, is not affected. The condition (\ref{found9}) does, however,
manifest itself in the form of ${\cal L}^{(4)}_{\eff}$: in the presence
of anomalies, this term is not gauge invariant. Wess and Zumino
\cite{Wess} have explicitly constructed an effective Lagrangian for which
the action transforms according to (\ref{found9}). The difference $\bar{{\cal
L}}^{(4)}_{\eff}={\cal L}^{(4)}_{\eff}-
{\cal L}_{\mbox{\scriptsize WZ}}$, therefore, yields a gauge
invariant action. The invariance theorem thus insures that, for a
suitable choice of the field variables, the quantity
$\bar{{\cal L}}^{(4)}_{\eff}=\bar{{\cal L}}^{(4)}_{\eff}[\pi,f]$
is invariant under the canonical transformation of
the fields $\pi,f$.

At higher orders of the expansion, the Wess-Zumino term also occurs in loop
graphs.
The corresponding contributions to the path integral are analyzed in detail in
the literature \cite{Akhoury}. It turns out that the Wess-Zumino term produces
a
noninvariant contribution to the generating functional exclusively through the
tree graphs of order $O(p^4)$ --- the loops yield gauge
invariant contributions.
This implies that the Lagrangian
\be\label{anom10}
{\cal L}_{\eff}[\pi,f]=\bar{{\cal L}}_{\eff}[\pi,f] +
{\cal L}_{\mbox{\scriptsize WZ}}[\pi,f] \ee
yields a generating functional obeying (\ref{found9}) if and only if the
contribution to the action from $\bar{{\cal L}}_{\eff}[\pi,f]$ is gauge
invariant. It thus suffices to equip the effective Lagrangian with the
appropriate Wess-Zumino
term --- the remainder has the same properties as if the theory were anomaly
free. Accordingly, the invariance theorem implies gauge invariance
of $\bar{{\cal L}}_{\eff}[\pi,f]$ to all orders of the derivative expansion.

\section{Approximate symmetries}\label{appr}\setcounter{equation}{0}

In the case of an approximate symmetry, the Lagrangian of the
underlying theory contains terms which explicitly break gauge invariance.
In the low energy domain, the
consequences of the
symmetry breaking are determined by the transformation
properties of the corresponding terms in the Lagrangian,
\be\label{appr1}
{\cal L}={\cal L}_0 +m_\alpha O^\alpha\fs\ee
The first term is invariant, while the operators
$O^\alpha$ transform with a nontrivial representation
$\hat{D}^\alpha_{\;\beta}(g)$ of G and the constants $m_\alpha$ determine the
strength of the symmetry breaking. In the case
of QCD, e.g., the breaking is bilinear in the quark fields, $O^\alpha=
(\,\bar{q}_{\mbox{\scriptsize R}}^{\;i}\,q_{\mbox{\scriptsize L}}^{\;j},\,
\bar{q}_{\mbox{\scriptsize L}}^{\;i}\,q_{\mbox{\scriptsize R}}^{\;j}\,)$, and
the elements of the quark mass matrix play the role of the symmetry breaking
parameters $m_\alpha$.

Since the
currents are not conserved, the generating functional considered in the
preceding sections fails to be
invariant under gauge transformations of the external fields ---
the Ward identities contain
additional contributions, generated by
the symmetry breaking part of the Lagrangian. These contributions involve Green
functions which not only contain the currents, but in addition involve the
operators $O^\alpha$. It is useful to extend the generating functional
accordingly, treating
the symmetry breaking parameters also as external fields,
on the same footing as
the vector
fields associated with the currents. The extended generating
functional
then contains two arguments, $\Gamma=\Gamma\{f,m\}$. The Green
functions of the operators $J^\mu_i,O^\alpha$ are obtained
by expanding this object in terms of the external fields $f_\mu^i(x)$ and
$m_\alpha(x)$. Note that, if the field $m_\alpha(x)$ is turned off, one is
dealing with
the symmetric theory, characterized by ${\cal L}_0$. To obtain the Green
functions in the presence of explicit symmetry breaking, the expansion
is to be performed around the nonzero, constant value of $m_\alpha$ which
occurs in ${\cal L}$.

In the absence of anomalies, the Ward identities are again equivalent to gauge
invariance of the extended generating functional. The only modification brought
about by the symmetry breaking terms is that the corresponding external fields
also transform under the action of the group. The transformation law involves
the representation carried by the operators $O^\alpha$:
\be\label{appr2}
T(g)m_\alpha=\hat{D}^\beta_{\;\alpha}(g^{-1})m_\beta\ee
The generating functional is invariant under a simultaneous transformation
of the two arguments,
\be\label{appr3}
\Gamma\{T(g)f,T(g)m\}=\Gamma\{f,m\} \fs\ee
(If anomalies occur, this relation is to be replaced by equation
(\ref{found9}) --- the form of the anomalous contributions
is not affected by the symmetry breaking, provided the dimension of the
operators $O^\alpha$ is smaller than the dimension of space-time.)

The analysis of the condition (\ref{appr3}) proceeds along the same
lines as before. The effective Lagrangian now involves two sets of external
fields
rather than one, ${\cal L}_{\eff}= {\cal L}_{\eff}[\pi,f,m]$ and the
derivative expansion now
also involves powers of the field $m_\alpha$.
The leading term is of the form $m_\alpha e^\alpha(\pi)$ --- it
is linear in $m_\alpha$ and does not contain derivatives of the pion field.
As it is the case
with the analogous quantities $g_{ab}(\pi),h^a_{\;i}(\pi)$, which specify
the leading contribution in the symmetric part of the effective
Lagrangian, gauge invariance fixes the form of the function $e^\alpha(\pi)$ in
terms of its values at $\pi=0$. The effective coupling constants $e^\alpha(0)$
are related to the
vacuum expectation values of the operators $O^\alpha$, which represent order
parameters of the spontaneously broken symmetry. The
number of independent coupling constants
permitted
by the continous part of the symmetry is equal to the number of one-dimensional
invariant subspaces of $D^\alpha_{\;\beta}(h),\;h\in\mbox{H}$; discrete
symmetries may impose additional constraints.

The Taylor expansion of the term $m_\alpha e^\alpha(\pi)$ in general contains a
contribution
which is quadratic in the pion fields, i.e., a pion mass term, with
$M_\pi^2\propto m_\alpha$. It is convenient to order the derivative expansion
accordingly, treating $m_\alpha$ as a quantity of order $p^2$. Needless to say
that an analysis in terms of effective fields is useful only if the symmetry
breaking parameters are sufficiently small, such that the pions remain light
and the pion pole dominance hypothesis still makes sense.

The inductive argument given in section \ref{high} goes
through without significant modifications,
because the specific transformation properties of the external fields do
not
play an important role in this context.
With a suitable change of variables, the action of the
effective theory may again be brought to a form where it is invariant under a
canonical gauge transformation of the arguments,
\be\label{appr4}
S_{\eff}\{\varphi(g,\pi),T(g)f,T(g)m\}=S_{\eff}\{\pi,f,m\}\fs\ee
This condition implies that the variational derivative
\be\label{appr5}
M^\alpha[\pi,f,m]=\frac{\delta S_{\eff}\{\pi,f,m\}}{\delta
m_\alpha(x)}\ee
transforms covariantly,
\be\label{appr6}
M^\alpha[\varphi(g,\pi),T(g)f,T(g)m]=D^\alpha_{\;\beta}(g)
M^\beta[\pi,f,m]\fs\ee
Integrating the quantity $\frac{d}{dt}S_{\eff}\{\pi,f,tm\}$
from $t=0$ to $t=1$, this yields
\be\label{appr7}
S_{\eff}\{\pi,f,m\}=S_{\eff}\{\pi,f,0\}+\int\!d^d\!x\int_0^1\!\!dt
m_\alpha M^\alpha[\pi,f,tm]\co\ee
such that the Lagrangian may be identified with
\be\label{appr8}
{\cal L}_{\eff}[\pi,f,m]={\cal L}_{\eff}[\pi,f,0]
+\int_0^1\!\!dtm_\alpha M^\alpha[\pi,f,tm] \fs\ee
In view of (\ref{appr6}), this convention insures that the part of the
Lagrangian which depends
on the field $m_\alpha(x)$ is manifestly gauge invariant. The remainder is
the
Lagrangian of the symmetric theory, where the preceding analysis applies as it
stands.

This shows that the invariance theorem also holds if the Lagrangian
of the underlying theory contains symmetry breaking terms.
In the framework of the effective theory, the
symmetry breaking parameters $m_\alpha$ act like spurions,
transforming contragrediently to the operators $O^\alpha$ which generate the
asymmetries.

\section{Summary and conclusion}\setcounter{equation}{0}
According to the Goldstone theorem, the spontaneous breakdown of a
continuous symmetry gives rise to massless particles, pions. The pion
pole dominance hypo\-the\-sis implies that the poles generated by the
exchange of these particles dominate the low energy structure of the theory.
Clustering then
requires that multipion exchange necessarily also occurs, generating cuts.
As pointed out in the early work on the subject,
the poles and cuts due to the Goldstone bosons may be
described in terms of an effective field theory, involving pion fields
as dynamical variables.

The
path integral formula (\ref{eff7}) provides the link between the underlying and
effective theories; it represents the
generating functional $\Gamma\{f\}$ of the Green functions formed
with the current operators in terms of an
effective Lagrangian. The derivation of this formula is based on general
kinematics and does not involve assumptions beyond the pion pole dominance
hypothesis.
The underlying theory does not fully determine
the effective Lagrangian, however:
\begin{enumerate}\item The operation ${\cal L}_{\eff} \rightarrow
{\cal L}_{\eff}+\partial_\mu\omega^\mu$ does not change the content of the
effective theory.
\item The pion fields
represent mere variables of integration. The effective theory remains the same
if the pion field is subject to a point transformation.
\end{enumerate}
In view of these ambiguities, which are inherent in the notion of an
effective Lagrangian, it is not evident that
the symmetries of the underlying theory insure a symmetric effective
Lagrangian. For the invariance of the path integral, it suffices that the
action is invariant --- the Lagrangian may pick up a total derivative.

Previous work on chiral perturbation theory is based on the assumption
that the effective Lagrangian does inherit the symmetry properties of the
underlying theory.
The assumption plays a crucial role in
the applications, because the symmetry is used
to determine the explicit form of the effective
Lagrangian.
The essence of the present paper is the statement that
the assumption is justified. The proof is rather involved, precisely because,
on account of the above ambiguities, the effective Lagrangian is partly a
matter of choice.

The proof exploits the fact that, in the absence of anomalies, the Ward
identities obeyed by the Green functions of the currents are equivalent to
gauge
invariance of the generating functional, i.e., to a local form of the symmetry.
The consequences for the effective Lagrangian are then
worked out by analyzing the perturbative expansion of the path integral.
The result is formulated as an invariance theorem,
which states that, in the absence of anomalies, the freedom of adding total
derivatives and performing a change of field variables may be used to bring the
effective Lagrangian to manifestly gauge invariant form. If the
underlying theory contains anomalies, the effective Lagrangian contains
a corresponding Wess-Zumino term --- the
remainder is gauge invariant.

The theorem establishes the relevant properties of the
effective Lagrangian as a consequence of the Ward identities and thus
puts chiral perturbation theory on a firm basis. Note
that the proof makes essential use of Lorentz invariance; the
theorem does not hold for nonrelativistic theories. The relevant
generalization is described elsewhere \cite{Ferromagnet}.
\newpage

{\LARGE {\bf Appendix}}
\appendix

\section{Construction of the map $\phi[g,\pi,f]$}\setcounter{equation}{0}
\label{map}
To establish the first one of the four assertions,
consider the difference
\be\label{ho10a}
\Delta_i[\pi,f]\equiv
D_\mu\frac{\delta S_{\eff}\{\pi,f\}_{n+1}}{\delta f^i_\mu (x)}-h^a_{\;i}(\pi)
\frac{\delta
S_{\eff}\{\pi,f\}_{n+1}}{\delta\pi^a(x)}
\co\ee
formed with the Killing vectors $h^a_{\;i}(\pi)$, which specify the
infinitesimal form of the canonical transformation law.
Since all of the terms except $S^{(n+1)}_{\eff}\{\pi,f\}$ are invariant under
the
canonical transformation of the fields $\pi$ and $f$, the function
$\Delta_i[\pi,f]$ only receives a contribution from this term, such that
$\Delta_i[\pi,f]=O(p^{n+1})$.

At the extremum of the classical action, $\Delta_i[\pi,f]$ vanishes. There, the
pion field is not an independent variable, but is subject
to the equation of motion, (\ref{ho10}).
In the present context,
this equation is needed only to leading
order, where it specifies the second
derivative of the pion field, $\ddot{\pi}$, in terms of
$\pi,\dot{\pi}$ and spacial derivatives thereof. The higher order time
derivatives may also be expressed in terms of these quantities.
At the extremum, the function $\Delta_i[\pi,f]$ thus
reduces to an expression which exclusively contains $\pi,\dot{\pi}$ and spacial
derivatives thereof. As discussed in section \ref{tree},
these are independent of
one another --- for the expression to vanish, it must vanish identically.

Next, dismiss the constraint on the pion field and consider the function
$\Delta_i[\pi,f]$ away from the extremum.
The higher order time derivatives may be eliminated in favour of the
variables $\pi,\dot{\pi}$ and their spacial derivatives, except that, instead
of a zero for the right hand side of the equation of motion,
the quantity $\delta S/\delta\pi $ and the derivatives thereof must now be
retained. Since the part
which does not contain these extra terms vanishes identically, the result is of
the form \be\label{ho18}
\Delta_i[\pi,f]=
\sum_{k=0}^{n-1}\eta^{a\;\mu_1\ldots\mu_k}_{\;i}[\pi,f]\partial_{\mu_1}\ldots
\partial_{\mu_k}
\frac{\delta S_{\eff}\{\pi,f\}_{n+1}}{\delta\pi^a(x)}
\fs\ee
This equation states that $\Delta_i[\pi,f]$ is the change occurring in
$S_{\eff}\{\pi,f\}_{n+1}$ under the shift
\be\label{ho19}
\delta\pi^a=
\bar{\eta}^a[g,\pi,f]\equiv
\sum_{k=0}^{n-1}(-1)^k\partial_{\mu_1}\ldots
\partial_{\mu_k}\left(g^i\,\eta^{a\;\mu_1\ldots\mu_k}_{\;i}[\pi,f]\right)\fs
\ee
of the pion field. Hence $S_{\eff}\{\pi,f\}_{n+1}$ is invariant under an
infinitesimal gauge transformation of the external fields, provided the pion
field is subject to the transformation
\be\label{ho19a}
\delta\pi^a=\;g^ih^a_{\,i}(\pi) +
\bar{\eta}^a[g,\pi,f]\fs
\ee
The statement holds for the action functional as such, not only at
the extremum. The local functions $\eta^{a\;\mu_1\ldots\mu_k}_{\;i}[\pi,f]$
represent a generalization of the Killing vectors $h^a_{\;i}(\pi)$.
Since
$\Delta_i[\pi,f]$ and $\delta S/\delta\pi $ are of order $n\!+\!1$ and 2,
repectively, the function $\bar{\eta}^a[g,\pi,f]$,
which specifies the modification of the transformation law, is a local
expression of order $n\!-\!1$.

Any finite
element $g^i(x)$ may be reached by a sequence of infinitesimal steps, e.g.,
along the path $g^i(x)_t=tg^i(x),\; 0\leq t\leq 1$. The transformation
law (\ref{ho19}) thus induces a mapping of the pion field also for finite
gauge transformations, which I denote by
$\pi^a\stackrel{g}{\rightarrow}\phi[g,\pi,f]$.
This verifies the first one of
the four properties listed in section \ref{it}: the
transformation of the pion field
just constructed insures
\be\label{inv10}
S_{\eff}\{\phi[g,\pi,f],T(g)f\}_{n+1}=S\{\pi,f\}_{n+1}+O(p^{n+2})\fs \ee

The map $\phi[g,\pi,f]$ deviates from
the canonical transformation $\varphi^a(g,\pi)$
only through the
contributions of order $n\!-\!1$, generated by $\bar{\eta}^a[g,\pi,f]$,
\be\label{ho20}
\phi^a[g,\pi,f]=\varphi^a(g,\pi)+\eta^a[g,\pi,f]
\fs\ee The local function $\eta^a[g,\pi,f]$ defined by this relation
generalizes
the quantity $\bar{\eta}^a[g,\pi,f]$ to group elements which are not in the
infinitesimal neighbourhood of unity.

\section{Composition law}
\label{cl}\setcounter{equation}{0}
To establish the composition law (\ref{inv7a}),
it is advantageous to express the higher order time derivatives occurring
in
$S^{(n+1)}_{\eff}\{\pi,f\}$ in terms of $\pi,\,\dot{\pi},\,\delta S/\delta \pi$
and the space derivatives thereof, in the manner discussed in appendix
\ref{map}. Note that the quantity $\delta S/\delta \pi$ is needed only to
leading order ---  the operation is exclusively applied to the highest order
term
of the truncated action. Integrating by parts, the result may be written in the
form
\be\label{inv11}
S^{(n+1)}_{\eff}\{\pi,f\}=\hat{S}^{(n+1)}_{\eff}\{\pi,f\}
-\int\!d^d\!x\psi^a[\pi,f]
\frac{\delta S}{\delta \pi^a(x)}\fs\ee
where $\hat{S}^{(n+1)}_{\eff}\{\pi,f\}$ only involves $\pi$ and $\dot{\pi}$,
while
$\psi^a[\pi,f]$ is a local function of order $p^{n-1}$.
The extra
term is equivalent to a shift in the pion field, i.e. to a change of variables,
\be\label{inv12}
\hat{\pi}^a =\pi^a +\psi^a[\pi,f]\fs\ee
Indeed, the
change of variables $\hat{{\cal L}}_{\eff}
[\hat{\pi},f]\equiv {\cal L}_{\eff}[\pi,f]$ leaves
the terms ${\cal L}^{(2)}_{\eff},\ldots,{\cal L}^{(n)}_{\eff}$
unaffected and modifies
the contribution of order $p^{n+1}$ in accordance with (\ref{inv11}). This
demonstrates that, with a suitable change of field variables,
$S^{(n+1)}_{\eff}\{\pi,f\}$ may be brought to a form where it involves the
pion field only through $\pi,\dot{\pi}$ and their space derivatives.

Adopting this choice of variables,
the quantity $\Delta_i[\pi,f]$ now contains at most two time derivatives of
the pion field; moreover, the expression is linear in $\ddot{\pi}$. In view of
(\ref{ho18}), this
immediately implies that the coefficients
$\eta^{a\;\mu_1\ldots\mu_k}_{\;i}[\pi,f]$ only contain $\pi$ and
$\dot{\pi}$ and, moreover, vanish if one of
the indices $\mu_1,\ldots,\mu_k$ is equal to zero. Accordingly,
the function $\eta[g,\pi,f]$, which specifies the transformation
law
of the pion field, only involves $\pi$ and $\dot{\pi}$ and the space
derivatives thereof. The same then holds true for the difference
$\phi[g_2,\phi[g_1,\pi,f],T(g_1)f]
- \phi[g_2g_1,\pi,f]$. In other words, the difference is the
same, irrespective of whether or not the pion field obeys the equation of
motion. Since the difference disappears when this equation is satisfied,
it vanishes identically.

As the change of variables used above singles out the time coordinate, it does
not preserve Lorentz invariance. It suffices, however, to
transform back to the ori\-gi\-nal coordinates --- the composition
law holds for all parametrizations of the pion field if it holds in one. This
verifies that the mapping
$\pi\stackrel{g}{\rightarrow}\phi[g,\pi,f]$ constructed in appendix \ref{map}
is a representation of the group.

\section{Canonical form of the transformation law}\setcounter{equation}{0}
\label{can}
The representation property of the map
$\pi\stackrel{g}{\rightarrow}\phi[g,\pi,f]$
amounts to a linear relation for the function $\eta[g,\pi,f]$,
\be\label{inv13}
\eta^a[g_2g_1,\pi,f]=\eta^a[g_2,\pi_1,T(g_1)f]
+\frac{\partial\varphi^a(g_2,\pi_1)}{\partial\pi^b_1}
\eta^b[g_1,\pi,f] \co\ee
with $\pi_1=\varphi(g_1,\pi)$. I first determine the general solution of this
relation and then
show that the solution differs from the trivial one, $\eta[g,\pi,f]=0$, only
by a change of variables.

The relation (\ref{inv13}), in particular, determines
the dependence of the function $\eta[g,\pi,f]$ on the pion field and its
derivatives: evaluation at $\pi=0$ yields a representation
in terms of the values at zero field.
The expression involves the representative group element
$n_\pi$ introduced in section \ref{it},
\be\label{inv14}
\eta^a[g,\pi,f]=
\eta^a[gn_\pi,0,T(n_\pi^{-1})f]-\frac{\partial\varphi^a(g,\pi)}{\partial\pi^b}
\eta^b[n_\pi,0,T(n_\pi^{-1})f]\ee
Furthermore, since the configuration $\pi=0$ is invariant under H,
the relation (\ref{inv13}) constrains the values of the
function at zero field,
\be\label{inv16}
\eta^a[gh,0,f]=\eta^a[g,0,T(h)f]+\varphi^a_{\;b}(g)
\eta^b[h,0,f]\;\;\;\;g\in\mbox{G},\,h\in\mbox{H}\ee
where $\varphi^a_{\;b}(g)$ is the derivative of the canonical map at
the origin,
\be\label{inv15}
\varphi^a_{\;b}(g)\equiv
\frac{\partial\varphi^a(g,\pi)}{\partial
\pi^b}\;\rule[-3mm]{.05mm}{8mm}_{\;\pi=0}\fs\ee
Differentiation of the composition law
$\varphi(g_2,\varphi(g_1,\pi))=\varphi(g_2g_1,\pi)$ shows
that the derivative of $\varphi(g,\pi)$ may be expressed in terms of the matrix
$\varphi^a_{\;b}(g)$, also for $\pi\neq 0$,
\be\label{inv15a}
\frac{\partial\varphi^a(g,\pi)}{\partial\pi^b}=
\varphi^a_{\;c}(gn_\pi)\varphi^c_{\;b}(n_\pi)^{-1}\fs\ee
Moreover, the matrix $\varphi^a_{\;b}(g)$ obeys
the product rule
$\varphi^a_{\;b}(gh)=\varphi^a_{\;c}(g)\varphi^c_{\;b}(h)$, valid for
$g\in\mbox{G},\,h\in\mbox{H}$. In
particular, $\varphi^a_{\;b}(h)$ is a representation of the subgroup H.

The element $g n_\pi$, which occurs in the first
term on the right hand side of equation (\ref{inv14}), may be decomposed as
$gn_\pi =n_{\pi_1}h_1$. In view of (\ref{inv16}), the function
$\eta[g,\pi,f]$ is thus fixed by its values on the two subspaces
$g=n,\pi=0$ and $g=h,\pi=0$:
\bea\label{inv17}
\eta^a[g,\pi,f]&\hspace{-1em}=&\hspace{-1em}
\eta^a[n_{\pi_1},0,T(n_{\pi_1}^{-1}g)f]
-\frac{\partial\varphi^a(g,\pi)}{\partial\pi^b}
\eta^b[n_\pi,0,T(n_\pi^{-1})f]\no
&\hspace{-0.2em}+&\hspace{-0.5em}\varphi^a_{\;b}(n_{\pi_1})
\eta^b[h_1,0,T(n_\pi^{-1})f] \fs\eea
This representation satisfies
the composition law (\ref{inv13}), provided the function $\eta[h,0,f]$
obeys the condition $(h,h^{\prime}\in\mbox{H})$
\be\label{inv21}
\eta^a[h,0,f]=\eta^a[hh^{\prime},0,T(h^{\prime\, -1})f]-
\varphi^a_{\;b}(h)\eta^b[h^{\prime},0,T(h^{\prime\, -1})f])\fs\ee
Note that, on the other subspace, the values are arbitrary --- the
function $\eta[n,0,f]$ is not subject to constraints. This is related to the
freedom in performing a transformation of the field variables (see
below).

The general solution of equation
(\ref{inv21}) may be obtained as follows. The relation connects
external fields which only differ by a gauge
transformation of the subgroup H. One may thus impose a gauge condition on
the external fields and use the relation to calculate the values of the
function
$\eta[h,0,f]$ for an arbitrary configuration in terms of
those on the subspace chosen by the gauge
condition.

A suitable
gauge condition is the following. Consider
those components of the vector
field $f^i_\mu(x)$ which correspond to the currents of the subgroup H.
At a given point of space-time, there exists a gauge, in which these components
of the field vanish,
together with the totally symmetric part of all of its derivatives,
\be\label{inv22}
\partial_{(\mu_1\ldots\mu_{k}}f^i_{\mu_{k+1})}(x)
= 0\;\;\;\;\;k=0,1,2,\ldots\ee
The condition fixes the gauge
uniquely, up to space-time independent transformations.

Constant gauge transformations may be disposed of as follows.
Consider equation (\ref{inv21}) with $h^\prime=h_0=\mbox{const.}$ and take
the average, integrating over $h_0\in\mbox{H}$.
This leads to the representation
\bea\label{inv23}
\eta^a[h,0,f]&\hspace{-0.5em}=&\hspace{-0.5em}\alpha^a[h,f]-
\varphi^a_{\;b}(h)\alpha^b[e,f]\no
\alpha^a[h,f]&\hspace{-0.5em}\equiv&\hspace{-0.5em}\int_{\mbox{\scriptsize H}}
d\mu(h_0)\eta^a[hh_0,0,T(h_0^{-1})f]\fs\eea
The volume of integration is the Haar
measure $d\mu(h_0)$, normalized to $\int_{\mbox{\scriptsize H}}
d\mu=1$, and $e$
stands for the unit element of the group.
Inserting the above representation in (\ref{inv21}), one obtains
\be\label{inv24}
\alpha^a[h,T(h^\prime)f]=\alpha^a[hh^\prime,f]-\varphi^a_{\;b}(h)
\alpha^b[h^\prime,f]+ \varphi^a_{\;b}(h)
\alpha^b[e,T(h^\prime)f]\fs\ee
The point is that, in view of the invariance of the measure,
the function $\alpha^a[h,f]$ is invariant under constant gauge
transformations, $\alpha[hh_0,T(h_0^{-1})f]=\alpha[h,f]$ ---
the equation to be solved is brought to a form where constant gauge
transformations are under control.

Suppose now
that $f$ is an arbitrary
configuration of the external field. A
suitable gauge transformation $T(h)$
takes it into the gauge (\ref{inv22}). Denote the field in this gauge
by $\bar{f}$, such that $f=T(h)\bar{f}$. It is essential that the
transformation $T(h)$ is local. The discussion concerns a fixed point of
space-time and the gauge condition is imposed only there. The transformation is
determined by the values of the gauge
field and its derivatives at that point, up to a constant.

Next, consider the quantity $\beta\equiv\alpha[h,\bar{f}]$.
In view of the invariance of $\alpha[h,f]$ under constant gauge
transformations,
$\beta$ only depends on the combination $f=T(h)\bar{f}$,
\be\label{inv25}
\alpha^a[h,\bar{f}]=\beta^a[T(h)\bar{f}]\fs\ee
The relation (\ref{inv24})
then immediately implies the more general representation
\be\label{inv26}
\alpha^a[h,f]=\beta^a[T(h)f]-\varphi^a_{\;b}(h)\beta^b[f]+
\varphi^a_{\;b}(h)\alpha^b[e,f])\co
\ee which also holds if $f$ does not obey the above gauge condition.
The corresponding representation for $\eta[h,0,f]$ takes the simple form
\be\label{inv27}
\eta^a[h,0,f]=\beta^a[T(h)f]-\varphi^a_{\;b}(h)\beta^b[f]\fs\ee
Indeed, one readily checks that this representation satisfies the constraint
(\ref{inv21}), irrespective of the form of $\beta[f]$.

The general solution of the representation property
(\ref{inv13}) is, therefore, of the form
\be\label{inv28}
\eta^a[g,\pi,f]=
\gamma^a[\pi_1,T(g)f] -\frac{\partial\varphi^a(g,\pi)}{\partial\pi^b}
\gamma^b[\pi,f]\co\ee
where the function $\gamma[\pi,f]$ receives a contribution, both from
$\eta[n,0,f]$ and from $\eta[h,0,f]$:
\be\label{inv29}
\gamma^a[\pi,f]=\eta^a[n_\pi,0,T(n_\pi^{-1})f] +\varphi^a_{\;b}(n_\pi)
\beta^b[T(n_\pi^{-1})f] \fs\ee

The result admits a very simple interpretation.
Consider a change of variables of the type
$\hat{\pi}=\pi +\psi[\pi,f]$. The corresponding change in the coordinates of
the transformed field $\phi[g,\pi,f]$ is given by
$\hat{\phi}=\phi+\psi[\phi,T(g)f]$. The operation thus modifies the form of the
transformation
function according to:
\be\label{inv30}
\hat{\eta}^a[g,\pi,f]=
\eta^a[g,\pi,f]+
\psi^a[\pi_1,T(g)f] -\frac{\partial\varphi^a(g,\pi)}{\partial\pi^b}
\psi^b[\pi,f]  \fs\ee
This is precisely of the form found from the solution of the
representation property. It thus suffices to perform the change of
variables
\be\label{inv31}
\psi^a[\pi,f]=-\gamma^a[\pi,f]\fs\ee
In the new coordinates, the transformation
law of the pion field takes the canonical form
$\pi\stackrel{g}{\rightarrow}\varphi(g,\pi)$, such that the action functional
obeys
(\ref{inv7d}) up to and including $O(p^{n+1})$.

\section{Gauge invariance of the Lagrangian}\setcounter{equation}{0}
\label{gau}
In this appendix, I determine the general form of the Lagrangian of a gauge
field theory, denoting the field and the group by $v_\mu(x)$ and
H, respectively. The Lagrangian ${\cal L}[v]$ is assumed to admit an expansion
in powers of the gauge field and its derivatives. The
essential ingredient of the analysis is the requirement that
the action functional $S\{v\}=\!\int\!d^d\!x\,{\cal L}[v]$ is gauge
invariant.

Although the result to be established is very simple, my
derivation, unfortunately, is rather clumsy. I work with the
variational derivative
\be\label{inv45} V^\mu
[v]=\frac{\delta S\{v\}}
{\delta v_\mu (x)}\fs\ee
Gauge invariance of the action implies that this quantity transforms
covariantly, \be\label{inv45a}
V^\mu[T(h)v]
=D(h)V^\mu[v]D(h^{-1})\ee
and obeys
\be\label{inv46}
D_\mu V^\mu[v]\equiv\partial_\mu V^\mu[v]-i[v_\mu,V^\mu[v]]
 =0\fs\ee
In the following, I solve these conditions and then study the
implications for the structure of the Lagrangian. The derivative
expansion may be ordered
in the standard manner, counting the field $v_\mu(x)$ and the derivative
$\partial_\mu$ as quantities of the same order. Since the transformation law
(\ref{high5}) of the gauge field
preserves the order, one may analyze the expansion term by
term, i.e., assume that the Lagrangian under consideration represents a
polynomial formed with the gauge field and its derivatives.

\subsection{Abelian gauge fields in $d=3$}
As a first step,
consider the case of an {\it abelian} group, denoting the components of the
gauge field by $v^i_\mu(x),\;i=1,\ldots,d_{\mbox{\scriptsize H}}$.
The transformation law
(\ref{inv45a}) then states that the local function $V_i^\mu[v]$ is gauge
invariant, $V_i^\mu[v+\partial h]=V_i^\mu[v]$. In this case, the
constraint (\ref{inv46})
is readily solved:
\be\label{inv47}
V_i^\mu[v]=\partial_\nu
K_i^{\mu\nu}[v]\;\;\;\;\;,\;\;\;\;\;K_i^{\mu\nu}[v]=-K_i^{\nu\mu}[v]\fs\ee

It is important here that the "potential" $K_i^{\mu\nu}$ is a local function,
i.e. only depends on the gauge field and its derivatives at one and the same
point of space-time.
Indeed, a more general version of
this statement is needed, valid for differential forms
\[\omega=\omega_{\mu_1\mu_2\ldots\mu_n}(x)\,
\mbox{d}x^{\mu_1}\wedge\mbox{d}x^{\mu_2}\wedge\ldots\wedge\mbox{d}x^{\mu_n}\,
,\] whose coefficients are local functions of the gauge field and its
derivatives, \[\omega_{\mu_1\mu_2\ldots\mu_n}(x)=\omega_{\mu_1\mu_2\ldots\mu_n}
\mbox{\Large(}v(x),
\partial v(x),\ldots\mbox{\Large)}. \] I
refer to these as {\it local} differential forms and indicate the argument in
the same manner as for ordinary fields: $\omega=\omega[v]$. The
relevant statement reads ($0\!<\!n\!<\!d$): If
$\omega[v]$ is a local $n$ -- form which obeys
$\mbox{d}\wedge\omega[v]=0$, then
there exists a local $(n-1)$ -- form $\Omega[v]$, such that
$\omega[v]=\mbox{d}
\wedge\Omega[v]$. Since this property is essential,
an explicit demonstration is given in appendix \ref{forms}.

The result immediately applies to the abelian form of the conservation law
(\ref{inv46}): the divergence of a vector field may be viewed as the exterior
derivative of a $(d\!-\!1)$ -- form. Hence the current $V_i^\mu[v]$ is the
exterior derivative of a local $(d\!-\!2)$ -- form, as claimed in
(\ref{inv47}).

Next, consider the transformation properties of the potential under a gauge
transformation.
Since the current is gauge invariant, the potential
$K^{\mu\nu}_i[v+\partial h]$ gives rise to the same current as
$K^{\mu\nu}_i [v]$, such that
$\partial_\mu
(K^{\mu\nu}_i[v+\partial h]-K^{\mu\nu}_i[v])=0$. One is thus again dealing
with a closed local differential form. According to appendix \ref{forms},
the difference may be represented as exterior
derivative of a local potential. In three dimensions, this yields
\be\label{inv48}
K^{\mu\nu}_i[v+\partial h]-K^{\mu\nu}_i[v]=\epsilon^{\mu\nu\rho}
\partial_\rho L_i[v,h]\co\ee
The relation implies that
the combination
\be\label{inv48a}
L_i[v,h_1+h_2]-L_i[v+\partial h_1,h_2] -L_i[v, h_1]=M_i\ee is a constant.
Since only the derivative of $L_i[v,h]$ matters, this function may be replaced
by $L_i[v,h]+M_i$. The above combination then vanishes,
so that, without loss of generality, one may set $M_i=0$.

In view of (\ref{inv48})
the gradient of $L_i[v,h]$ does not depend on the value of $h$, but
only on the derivatives thereof. Hence the expression itself contains $h$ at
most linearly,
$L_i[v,h]=c_{ik}h^k+\bar{L}_i$, where the coefficients $c_{ik}$ and $\bar{L}_i$
are local
functions of $v$ and $\partial h$. Actually, for the quantity $h$ to disappear
upon taking the gradient, the coefficient $c_{ik}$ must be a constant, such
that
\be\label{inv49}
L_i[v,h]=c_{ik}h^k+\bar{L}_i[v,\partial h]\fs\ee
The corresponding decompositions of the function $K^{\mu\nu}_i[v]$ and of the
Lagrangian take the form \be\label{inv49a}
K^{\mu\nu}_i[v]=
c_{ik}\epsilon^{\mu\nu\rho}v_\rho^k+\bar{K}^{\mu\nu}_i[v]\;\;\;\;,\;\;\;\;
{\cal L}[v]=
{\cal L}_{\mbox{\scriptsize CS}}[v]+\bar{{\cal L}}[v]\co\ee
where ${\cal L}_{\mbox{\scriptsize CS}}[v]$
is the abelian version of the Chern-Simons Lagrangian,
\be\label{inv50}
{\cal L}_{\mbox{\scriptsize CS}}[v]=\mbox{$\frac{1}{2}$}
c_{ik}\epsilon^{\mu\nu\rho}v^i_\mu\partial_\nu v_\rho^k\fs\ee
Although this expression fails to be gauge invariant, the corresponding action
is invariant under gauge transformations (recall that only
"small" external fields are relevant, which may be taken to
vanish outside some finite region of space-time).

Next, consider the function $\bar{L}_i[v,\partial h]$.
Since the quantity $M_i$ vanishes, this function obeys the condition
\be\label{inv51}
\bar{L}_i[v,\partial h_1+\partial h_2]-\bar{L}_i[v+\partial h_1,\partial h_2]-
\bar{L}_i[v, \partial h_1]=0\co\ee
which may be solved with the technique used in appendix \ref{can}.
The gauge field admits the unique decomposition
$v^i_\mu=\bar{v}^i_\mu +\partial_\mu h^i$, where $\bar{v}^i_\mu$ obeys
the gauge condition (\ref{inv22}). Accordingly, there is a local
function $N_i[v]$ such that
\be\label{inv52}
N_i[v]=\bar{L}_i[\bar{v},\partial h]\fs\ee The composition law
(\ref{inv51}) then entails the representation
\be\label{inv53}
\bar{L}_i[v,\partial h]=N_i[v+\partial h]-N_i[v]\co\ee
valid $\forall \,v,h$. Finally, the relation (\ref{inv48}) shows that the
quantity
\[\tilde{K}^{\mu\nu}_i[v]=\bar{K}^{\mu\nu}_i[v]-
\epsilon^{\mu\nu\rho}\partial_\rho N_i[v]\] is gauge invariant. Now,
$\bar{K}^{\mu\nu}_i[v]$ may be replaced by
$\tilde{K}^{\mu\nu}_i[v]$, without changing the current $V^\mu_i[v]$. Once
the
Chern-Simons term is removed, the variational derivative of the action may,
therefore, be represented in terms of a gauge invariant potential,
$\bar{V}^\mu_i[v]=\partial_\nu\tilde{K}^{\mu\nu}_i[v]$. The
corresponding expression for the action is obtained by integrating along the
path $tv_\mu^i$ from $t=0$ to $t=1$,
\be\label{inv55}
\bar{S}\{v\}=
\int\!d^3\!x\int_0^1\!\!dtv^i_\mu\partial_\nu\tilde{K}^{\mu\nu}_i[tv]\fs\ee
An integration by parts in the second term leads to a gauge invariant
expression for the corresponding contribution to
the Lagrangian,
\be\label{inv56}
\bar{{\cal L}}[v]=
\int_0^1\!\!dt\partial_\mu v^i_\nu \tilde{K}^{\mu\nu}_i[tv]
\fs\ee

\subsection{Abelian gauge fields in $d=4$}
The calculation proceeds along the same lines also in four dimensions. The
relation (\ref{inv48}) now takes the form
\be\label{inv57}
K^{\mu\nu}_i[v+\partial h]-K^{\mu\nu}_i[v]=
\epsilon^{\mu\nu\rho\sigma}\partial_\rho
L_{i\sigma}[v,h]\co\ee
where $L_{i\sigma}[v,h]$ is a local function of its arguments,
determined by this equation up to a gradient. The left hand side only
involves the derivatives of $h$. I first show that, without loss of
generality, the function $L_{i\sigma}[v,h]$ may be taken to have the same
property.

As mentioned above, the discussion may be restricted to Lagrangians of
polynomial form.
Accordingly, it
suffices to analyze the properties of the function $K^{\mu\nu}_i[v]$ under the
assumption that one is dealing with a polynomial of the gauge field and its
derivatives. The left hand
side of (\ref{inv57}) then represents a polynomial in the variables $v$ and
$\partial h$.
Since the divergence of the expression vanishes identically, one may collect
terms with a given degree of homogeneity in $h$ and represent each of these
as a rotation, such that $L_{i\sigma}[v,h]$ takes the form of
a polynomial in the variable $h$ and its derivatives. Extracting those
factors which do not contain derivatives, the expression takes the form
$L_{i\sigma}[v,h]=\sum l_{i\sigma,i_1,\ldots
i_k}h^{i_1}\ldots h^{i_k}$, where the coefficients only involve $v$ and
$\partial h$. According to (\ref{inv57}), the rotation thereof does not contain
any such factors. This
implies, in particular, that the coefficients of the terms with the largest
value of $k$ are rotation free and may thus be written as a gradient.
Removing a suitable
gradient from $L_{i\sigma}[v,h]$, the largest
value of $k$ is reduced by one unit. Proceeding in this way until no factors of
$h$ are left,
one arrives at an expression of the form
$L_{i\sigma}=L_{i\sigma}[v,\partial h]$, thus verifying the above claim.

The relation (\ref{inv57}) implies
\be\label{inv59}
L_{i\sigma}[v, \partial h_1+\partial h_2]
-L_{i\sigma}[v+\partial h_1,\partial h_2]-
L_{i\sigma}[v,\partial h_1]=\partial_\sigma M_i[v,h_1,h_2]\fs\ee
This relation, in turn, requires the
combination
\be\label{inv60}
M_i[v,h_1+h_2,h_3]-M_i[v,h_1,h_2+h_3] -M_i[v+\partial h_1,h_2,h_3]
+M_i[v,h_1,h_2]\ee
to be a constant. As a local quantity, the expression only depends on the
values of the
fields and their derivatives at the point under consideration. It can only be
constant, if it is independent of these fields. Since
the combination vanishes for $h_i=0$, it vanishes altogether.

According to (\ref{inv59}), the gradient of
$M_i[v,h_1,h_2]$ only involves the derivatives of $h_1$ and $h_2$. Hence, the
variables themselves enter at most linearly,
\be\label{inv61} M_i[v,h_1,h_2]=
c^1_{ik}h_1^k +c^2_{ik}h^k_2 +\bar{M}_i[v,\partial h_1,\partial h_2]\co \ee
where $c^1_{ik}$ and $c^2_{ik}$ are constants. So, the
combination (\ref{inv60}) contains at most a term linear in
$h$, viz. $c^1_{ik}h^k_1-c^2_{ik}h^k_3$. This term, however, only vanishes
$\forall \;h_1,h_3$ if	$c^1_{ik}=c^2_{ik}=0$. Hence the function
$M_i[v,h_1,h_2]$ exclusively involves the derivatives of the arguments
$h_1,h_2$.

Invoking the decomposition $v^i_\mu=\bar{v}^i_\mu +
\partial_\mu h^i$, this property allows the construction of
a local quantity $N_i[v,\partial h]$, with
\be\label{inv70}
N_i[v,\partial h_1]=\bar{M}_i[\bar{v},\partial h,\partial h_1]\fs\ee The
composition rule (\ref{inv60}) then yields
\be\label{inv71}
M_i[v, h_1, h_2]=-N_i[v,\partial h_1+\partial h_2]
+N_i[v+\partial h_1,\partial h_2]+N_i[v,\partial h_1]\ee
$\forall \,v,h_1,h_2$. Hence the function $N_i[v,\partial h]$ may be absorbed
in $L_{i\sigma}[v,\partial h]$ --- without loss of generality, one may set
$M_i[v,h_1,h_2]=0$. Applying the same argument once more to the
composition rule (\ref{inv59}), one verifies that the function
$L_{i\sigma}[v,\partial h]$ admits a representation of the
form
\be\label{inv72}
L_{i\sigma}[v,\partial h]=P_{i\sigma}[v+\partial h]-P_{i\sigma}[v]\fs\ee
The relation (\ref{inv57}) then shows that the function
$P_{i\sigma}[v]$ may be absorbed in $K^{\mu\nu}_i[v]$. In this
convention, the quantity $L_{i\sigma}[v,\partial h]$ vanishes, such that
the potential $K^{\mu\nu}_i[v]$ becomes gauge invariant.
An analogue of the Chern-Simons Lagrangian does, therefore, not occur in
$d=4$. In the abelian case under discussion here, the Lagrangian may always be
brought to the manifestly gauge invariant
form
\be\label{inv73}
{\cal L}[v]=\int_0^1\!\!dt\partial_\mu v^i_\nu K^{\mu\nu}_i[tv]
\fs\ee

\subsection{Nonabelian gauge fields}
The extension of the above calculation to the nonabelian case runs as follows.
The Lagrangian consists of a
series of vertices of the type $\partial^Dv^E$,
where $D$ counts the overall number of derivatives and $E$ is the
number of gauge fields entering the term in question. For the present
purpose, it is convenient to order the vertices according to the number $E$ of
gauge fields and to use induction in the value of $E$.
For definiteness, I consider the three-dimensional case --- the
extension to $d=4$ is trivial.

Consider those vertices which contain the minimal number of gauge
fields,
$E=2$ and denote the corresponding contribution to the effective Lagrangian
by ${\cal L}[v]_2$.
The variational derivative of $\int\!d^3\!x {\cal
L}[v]_2$ yields a current of $O(v^1)$, which I call
$V^\mu_i [v]_1$.
Since the conservation law (\ref{inv46}) holds term by term in the above
counting of powers, it implies
\be\label{inv80}
\partial_\mu V^\mu_i[v]_1=0\fs\ee
All other contributions involve at least
two gauge fields.

Under an infinitesimal gauge transformation, $\delta v_\mu =\partial_\mu
h -i[v_\mu,h]$, the generic term of order
$\partial^Dv^{E-1}$, which occurs in the expansion of the current, yields
contributions of order $\partial^{D+1}v^{E-2}h$ as
well as terms of order $\partial^Dv^{E-1}h$. The former are produced by the
abelian gauge
transformation $v_\mu\rightarrow v_\mu+\partial_\mu h$, while the latter arise
from the operation
$v_\mu\rightarrow v_\mu-i[v_\mu,h]$. Comparing terms of order $\partial^Dv^0h$
on the two sides of the transformation law (\ref{inv45a}), one obtains
\be\label{inv81}
V^\mu_i[v+\partial h]_1=V^\mu_i[v]_1\fs\ee
In other words, the part of the Lagrangian which contains the smallest
number of gauge fields is symmetric under a group of abelian
transformations, i.e., obeys the same equations
as the full current in the abelian case.
In ${\cal L}[v]_2$, the nonabelian character of the group only
manifests itself through a supplementary condition: for
space-time
independent transformations, the transformation law (\ref{inv45a}) implies that
the quantity $V^\mu[v]_1=\sum_it_iV^\mu_i[v]_1$ contains the various abelian
fields in such a combination that the result transforms
covariantly,
\be\label{inv82}
V^\mu[T(h_0)v]_1=D(h_0)V^\mu[v]_1D(h_0^{-1})\fs\ee
The condition amounts to a constraint on the form of the possible
couplings --- it selects a subset of the Lagrangians permitted by
abelian symmetry.

The results of the preceding analysis may now be taken over as they are.
The Lagrangian in general contains a Chern-Simons term, ${\cal
L}[v]_2=
{\cal L}_{\mbox{\scriptsize CS}}[v]+\bar{{\cal L}}[v]_2$. The remainder,
$\bar{{\cal L}}[v]_2$, is invariant under
abelian gauge transformations.
The Chern-Simons term illustrates the
constraint mentioned above: Suppose that the group H is simple. The relation
(\ref{inv82}) then requires that the coefficients $c_{ik}$, which enter the
expression for ${\cal L}_{\mbox{\scriptsize CS}}[v]$, are determined by a
single coupling constant:
$\frac{1}{2}c_{ik}=c\mbox{tr}(t_it_k)$, while, for an abelian
theory, the symmetry does not constrain the values of these couplings.
Similar relations among the various independent coupling constants of the
abelian theory, naturally, also arise for the gauge invariant part of the
Lagrangian (note that $\bar{{\cal L}}[v]_2$ collects an infinity of
vertices, containing an arbitrary number of derivatives).

Since the term ${\cal L}[v]_2$ only represents the part
of the Lagrangian with the smallest number of gauge fields, the corresponding
action $\int\!d^3\!x{\cal L}[v]_2$ is invariant
only under abelian transformations. One may, however, add suitable
higher order terms to arrive at a fully gauge invariant result. In the case of
the Chern-Simons term, it suffices to  add the familiar contribution
of order $v^3$,
\be\label{inv83}
{\cal L}_{\mbox{\scriptsize CS}}[v]=
c\epsilon^{\lambda\mu\nu}
\mbox{tr}\{v_\lambda\partial_\mu v_\nu
-\mbox{$\frac{2}{3}$}iv_\lambda v_\mu v_\nu\}\ \fs\ee
The remainder, $\bar{{\cal L}}[v]_2$,
is invariant under
abelian gauge transformations and may, therefore, be expressed in terms of the
abelian field strength $\partial_\mu v_\nu-\partial_\nu v_\mu$ and the
derivatives thereof. To render the expression gauge invariant with respect to
H, it suffices to augment the abelian field strength by the standard
contribution involving the commutator $[v_\mu,v_\nu]$ and to replace the
derivatives of the field strength by covariant ones. In view of the fact that
the expression
is invariant under constant gauge transformations, it is automatically
invariant under the full gauge group --- the field strength and its
covariant derivatives transform homogeneously. This results in a
representation for the Lagrangian which correctly describes
the vertices of order $v^2$
and, moreover, yields a gauge invariant action.

Finally, consider the higher order terms. Removing the part of the action
just constructed, one remains with an expression which is gauge
invariant under H and only contains vertices of order $v^3$ or higher.
Collect the vertices of $O(v^3)$ in
${\cal L}[v]_3$ and repeat the above analysis.
There is a simplification in so far as an abelian
Chern-Simons term only occurs at $O(v^2)$. So, from
the second iteration on, the quantity ${\cal L}[v]_n$
is invariant under abelian gauge transformations.

The net result is an
expression for the effective Lagrangian which is gauge invariant under H,
except for the term ${\cal L}_{\mbox{\scriptsize CS}}$
specified above,
\be\label{inv84}
{\cal L}[v]=
{\cal L}_{\mbox{\scriptsize CS}}[v]+\bar{{\cal L}}[v]
\;\;\;\;\;,\;\;\;\;\; \bar{{\cal L}}[T(h)v]=\bar{{\cal L}}[v]
\fs\ee The same representation also holds for $d=4$, except that the term
${\cal L}_{\mbox{\scriptsize CS}}[v]$ is then absent. This completes the
derivation of the result stated at the beginning of the present appendix.

The application to the effective Lagrangian is straightforward. According to
section \ref{high}, the quantity $S_{\eff}\{0,v,0\}$ is gauge invariant under
H. The above result implies that the corresponding part of the effective
Lagrangian, ${\cal L}_{\eff}[0,v,0]$,
is gauge invariant, up to a possible Chern-Simons term. Putting things
together, one concludes that the full effective Lagrangian is gauge invariant,
except for the contribution from ${\cal L}_{\mbox{\scriptsize CS}}[v]$.
Note, however, that the field $v_\mu$ occurring therein does
not coincide with the original external field, but differs from it through a
gauge transformation, which depends on the pion field: according to
(\ref{inv40a}), the quantity to be inserted
in the Chern-Simons Lagrangian is the vector component of
$f_{\pi\,\mu}=T(n_\pi^{-1})f_\mu =v_\mu+a_\mu$.

To see how the pion field enters the result, consider the Chern-Simons
Lagrangian built with the whole field $f_\pi$,
\be\label{inv84a}
\bar{{\cal L}}_{\mbox{\scriptsize CS}}[f_\pi]=
c\epsilon^{\lambda\mu\nu}
\mbox{tr}\{f_{\pi\,\lambda}\partial_\mu f_{\pi\,\nu}
-\mbox{$\frac{2}{3}$}if_{\pi\,\lambda} f_{\pi\,\mu} f_{\pi\,\nu}\} \fs\ee
Inserting the decomposition $f_{\pi\,\mu}=v_\mu+a_\mu$, this
gives
\be\label{inv85}
\bar{{\cal L}}_{\mbox{\scriptsize CS}}[f_\pi]=
{\cal L}_{\mbox{\scriptsize CS}}[v ]+
c\epsilon^{\lambda\mu\nu}\mbox{tr}\{a_\lambda D_\mu a_\nu\}
-\mbox{$\frac{2}{3}$}ic\epsilon^{\lambda\mu\nu}\mbox{tr}\{a_\lambda a_\mu
a_\nu\}\co\ee
with $D_\mu a_\nu=\partial_\mu a_\nu -i[v_\mu,a_\nu]$. The point is that the
extra
terms represent tensorial contributions which are gauge invariant under H. One
may thus replace ${\cal L}_{\mbox{\scriptsize CS}}[v]$ by
$\bar{{\cal L}}_{\mbox{\scriptsize CS}}[f_\pi]$, compensating for the
difference in the remaining, gauge invariant part of the Lagrangian.

The dependence of $\bar{{\cal L}}_{\mbox{\scriptsize CS}}[f_\pi]$ on the pion
field is readily worked out. The field enters through the gauge
transformation $f_{\pi\,\mu} =D^{-1}f_\mu D+iD^{-1}\partial_\mu D$, with
$D=D(n_\pi)$. Using the abbreviation $\omega_\mu\equiv (-i)\partial_\mu D
D^{-1}$, this gives
\be\label{inv86}
\bar{{\cal L}}_{\mbox{\scriptsize CS}}[f_\pi]
=\bar{{\cal L}}_{\mbox{\scriptsize CS}}[f] +
c\epsilon^{\lambda\mu\nu}\partial_\lambda\mbox{tr}\{\omega_\mu f_\nu\}
-i\mbox{$\frac{1}{3}$}c\epsilon^{\lambda\mu\nu}\mbox{tr}\{\omega_\lambda
\omega_\mu\omega_\nu\}\fs\ee
Both the second and the third term represent total derivatives, and may thus
be discarded (for the third term, this can be shown, e.g., by calculating the
change produced by a variation of the pion field). Hence the field
$f_\pi$ may be replaced by the external field $f$ --- the
action generated by $\bar{{\cal L}}_{\mbox{\scriptsize CS}}[f_\pi]$
is independent of the pion field, as claimed in assertion {\bf D} of section
\ref{it}.

\section{Closed local differential forms}\setcounter{equation}{0}
\label{forms}

The proof given in appendix \ref{gau} makes essential use of the fact that
closed local forms may be expressed as derivatives of a potential which is
itself local. The derivation of this statement relies on an elementary property
of differential
forms: if the closed $n$ -- form $f$ ($0\!<\!n\!<\!d$) vanishes outside a
ball $V$, \[ \mbox{d}\wedge f(x) =0\;\;\mbox{and}\;\;f(x)=0\;\;\;\forall\,
x\not\in V\;,\] then it is the
exterior derivative of an $(n\!-\!1)$ -- form, {\it which also vanishes
outside} $V$:
\[f(x)=\mbox{d}\wedge F(x)\;\;\mbox{and}\;\;F(x)=0\;\;\;\forall\,
x\not\in V\;.\]
Presumably, this is a special case of a more general statement, valid, e.g.,
for simply connected regions.
As I did not find this mentioned in the standard textbooks, I present
an explicit demonstration for the case of a ball --- this
suffices for the present purposes.

If $f(x)$ is a one-form, $f(x)=f_\mu(x)\,\mbox{d}x^\mu$, the statement
immediately follows from the
explicit representation $F(x)=\int_a^xdy^\mu f_\mu(y)$: it suffices to choose
the starting point $a$ of the path of integration outside of $V$. Since
$\mbox{d}\wedge f$ vanishes, the
integral is independent of the path chosen to reach the point $x$. Hence, if
$x$ is outside $V$, one may take a path which does not enter $V$
at all, such
that $F(x)=0$, as claimed. For higher forms, the property may be established
by means of
induction. Assume that it holds for $(n\!-\!1)$ -- forms. Isolating one of
the coordinates, say $t\equiv x^d$, any $n$ -- form $f$ defined on a
$d$ -- dimensional manifold $M^d$ gives rise to
two forms $g_t$ and $h_t$ which live on the
$(d\!-\!1)$ -- dimensional manifold $M^{d-1}$ with coordinates
$\hat{x}=(x^1,x^2,\ldots,x^{d-1})$,
while the remaining variable, $t$, only enters parametrically,
\[ f(x)=\mbox{d}t\wedge g_t(\hat{x})+h_t(\hat{x})
\;\;\;;\]
$g_t(\hat{x})$ is an $(n\!-\!1)$ -- form, while $h_t(\hat{x})$ is an $n$
-- form. The vanishing of
$\mbox{d}\wedge f(x)$ entails two separate conditions on $g_t(\hat{x})$
and $h_t(\hat{x})$:\[\hat{\mbox{d}}\wedge
g_t(\hat{x})=\dot{h}_t(\hat{x})\;\;\;,\;\;\hat{\mbox{d}}\wedge
h_t(\hat{x})=0\co\]
where $\hat{\mbox{d}}$ is the exterior derivative on $M^{d-1}$
and the dot indicates a derivative with respect to the parameter $t$.
The integral\[
G_t(\hat{x})=\int_{\!-\infty}^t\!\!\!dt^{\prime}\;g_{t^\prime}(\hat{x})\]
obeys $\dot{G}_t(\hat{x})=g_t(\hat{x}),\; \hat{\mbox{d}}\wedge G_t(\hat{x})
= h_t(\hat{x})$ and hence represents a potential for $f$, $\mbox{d}\wedge
G_t(\hat{x})= f(x)$. It does not quite solve the problem, however, because
$G_t(\hat{x})$ does not necessarily vanish in the shadow $V^+$ cast by the
ball under
illumination along the
$t$ -- axis from below. There, the integral is
independent of $t$,
$G_t(\hat{x})=\bar{G}(\hat{x})$ and obeys
$\hat{\mbox{d}}\wedge\bar{G}(\hat{x})=0$. Since $\bar{G}(\hat{x})$ is an
$(n\!-\!1)$ -- form which vanishes outside the projection of the ball onto
$M^{d-1}$,
the induction hypothesis implies that there
is a form $\bar{H}(\hat{x})$, which also vanishes there and obeys
$\hat{\mbox{d}}\wedge\bar{H}(\hat{x})=\bar{G}(\hat{x})$. Now, take a smooth
function $\chi(\hat{x},t)$, which interpolates
between the value $1$ on $V^+$ and the value $0$ on the opposite
side, $V^-$, but is otherwise arbitrary (strictly speaking, to avoid singular
behaviour at the intersection of $V^+$ with $V^-$, one must enlarge the ball
slightly, cutting out a small shell from $V^+$ and $V^-$, such that the
intersection disappears; accordingly the construction only insures the
vanishing of the potential outside a region which is somewhat larger
than $V$). The form $H_t(\hat{x})=\chi(\hat{x},t)\bar{H}(\hat{x})$
obeys $\hat{\mbox{d}}\wedge H_t=G_t,\;\dot{H}_t=0 $ everywhere outside $V$.
Hence
the quantity\[ F(x)\equiv G_t(\hat{x})-\mbox{d}\wedge H_t(\hat{x})\]
vanishes outside $V$ and obeys $\mbox{d}\wedge F(x)= f(x)$; this verifies the
claim.

Next, consider a differential form $\omega[v]$, whose coefficents
only involve a set of fields $v$ and their derivatives at the given point of
the manifold (in the terminology used in appendix
\ref{gau}: a local differential form). Suppose that the form is closed,
$\mbox{d}\wedge \omega[v]
=0$. The claim is that there is a local differential form $\Omega[v]$, such
that $\mbox{d}\wedge \Omega[v]=\omega[v]$. To verify this, consider a
deformation $\delta v$ of the fields. The corresponding
change in $\omega[v]$ is of the form
$\delta\omega[v]=\vartheta[v]\cdot\delta v$, where $\vartheta[v]$ is a
differential operator, whose
coefficients are local functions of the fields. The operator
obeys $\mbox{d}\wedge \vartheta[v]=0$. Application of the above construction
shows that there is a kernel $K_v(x,y)$ which (i) satisfies
$\mbox{d}\wedge K_v(x,y)=\vartheta[v]\cdot\delta(x,y)$ and (ii) vanishes
outside the region where $\vartheta[v]\cdot\delta(x,y)$ is different from zero.
In other words, the support of the kernel is the point $x=y$, such that
$K_v(x,y)$ may be represented in terms of a local differential operator
$\theta[v]$ acting on the $\delta$-function,
$K_v(x,y)=\theta[v]\cdot\delta(x,y)$. The operator obeys
$\mbox{d}\wedge\theta[v]=\vartheta[v]$. Accordingly, the form $\delta\omega[v]$
admits a local potential, $\delta\Omega=\int\!d^d\!yK_v(x,y)\delta
v(y)=\theta[v]\cdot\delta v$. Finally, this expression may be integrated along
the
path $tv(x)\,,\,0\leq t\leq1$, which corresponds to a sequence of deformations,
$\delta v(x)=dt\,v(x)$. The quantity
$\Omega[v]=\int_0^1\!dt\;\theta[tv]\cdot v$ is a local form which obeys
$\mbox{d}\wedge \Omega[v]=\omega[v]$. This verifies the statement used in
appendix \ref{gau}. 
\end{document}